\documentclass[preprint,3p,twocolumn,numbers,sort&compress]{elsarticle}

\usepackage{amssymb}
\usepackage{amsmath}
\usepackage{dsfont}
\usepackage{bm,bbm}
\usepackage{lipsum}
\usepackage{mathtools, cuted}
\usepackage{breqn}
\usepackage{xcolor}
\usepackage{soul}
\usepackage{caption}
\usepackage{subcaption}
\usepackage{graphicx}

\usepackage{amsthm}
\usepackage{verbatim}
\usepackage{hyperref}

\newtheorem{thm}{Theorem}
\newtheorem{cor}{Corollary}

\journal{Journal of Economic Behavior \& Organization}

\begin{document}

\begin{frontmatter}

\title{Correlation between upstreamness and downstreamness in random global value chains}

\author[inst1,inst2]{Silvia Bartolucci}

\affiliation[inst1]{organization={Dept. of Computer Science, University College London},
            addressline={66-72 Gower Street}, 
            city={London},
            postcode={WC1E 6EA}, 
            country={United Kingdom}}   
\affiliation[inst2]{organization={Centre for Financial Technology, Imperial College Business School},
            addressline={South Kensington}, 
            city={London},
            postcode={SW7 2AZ}, 
            country={United Kingdom}}

\author[inst1,inst3,inst4]{Fabio Caccioli}

\affiliation[inst3]{organization={Systemic Risk Centre, London School of Economics and Political Sciences},
            addressline={}, 
            city={London},
            postcode={WC2A 2AE}, 
            country={United Kingdom}}
            
\affiliation[inst4]{organization={London Mathematical Laboratory},
addressline={8 Margravine Gardens}, 
            city={London},
            postcode={WC 8RH}, 
            country={United Kingdom}}

\author[inst5]{Francesco Caravelli}

\affiliation[inst5]{organization={Theoretical Division (T4), Condensed Matter $\&$ Complex Systems,
Los Alamos National Laboratory},
            city={Los Alamos},
            postcode={87545}, 
            state={New Mexico},
            country={U.S.A.}}

\author[inst6]{Pierpaolo Vivo}

\affiliation[inst6]{organization={Dept. of Mathematics, King’s College London},
            addressline={Strand}, 
            city={London},
            postcode={WC2R 2LS}, 
            country={United Kingdom}}

\begin{abstract}
This paper is concerned with upstreamness and downstreamness of industries and countries in global value chains. Upstreamness and downstreamness measure respectively the average distance of an industrial sector from final consumption and from primary inputs, and they are computed from
the most used global Input-Output tables databases, e.g., the World Input-Output Database (WIOD). Recently, Antr\`as and Chor reported a puzzling and counter-intuitive finding in data from the period 1995-2011, namely that (at country level) upstreamness appears to be positively correlated with downstreamness, with a correlation slope close to $+1$. This effect is stable over time and across countries, and it has been confirmed and validated by later analyses. We first analyze a simple model of random Input/Output tables, and we show that, under minimal and realistic structural assumptions, there is a natural positive correlation emerging between upstreamness and downstreamness of the same industrial sector/country, with correlation slope equal to $+1$. This effect is robust against changes in the randomness of the entries of the I/O table and different aggregation protocols. Secondly, we perform experiments by randomly reshuffling the entries of the empirical I/O table where these puzzling correlations are detected, in such a way that the global structural constraints are preserved. Again, we find that the upstreamness and downstreamness of the same industrial sector/country are positively correlated with slope close to $+1$, even though the random reshuffling has destroyed any underlying economic information about inter-sectorial connections and trends. Our results -- rooted in the Complexity Science approach to economic problems -- strongly suggest that \textcolor{black}{(i) extra care is needed when interpreting these measures as simple representations of each sector's positioning along the value chain, as the ``curse of the \textcolor{black}{input-output identities}'' and labor effects effectively force the value chain to acquire additional links from primary factors of production, and (ii)} the empirically observed puzzling correlation may rather be a necessary consequence of the few structural constraints (positive entries, and sub-stochasticity) that Input/Output tables and their surrogates must meet\textcolor{black}{, in turn making other proposed measures of sector inter-linkages more suitable and intuitive.}
\end{abstract}


\end{frontmatter}

\section{Introduction}
\label{sec:Introduction}

The structure of national and international trade flows has undergone a dramatic transformation in the past decades. Understanding how global value chains shape the exchange of goods and money at different scales (from industrial sectors to countries) has become of central importance. Researches on these issues usually rely on Input-Output analysis -- the field pioneered by V. Leontief \cite{Leontiefbook, Leontief1936}. This level of analysis is facilitated by the increasing availability and development of detailed Input/Output (I-O) tables for each country \cite{handbook,wiotdataset}.

To characterize the complexity of global value chains, metrics have been devised that take such empirical I-O tables as starting point. In particular, Antr\`as and Chor  \cite{antras2012econometrica}, Miller and Temurshoev \cite{Miller} and Fally \cite{Fally2012} introduced the notions of \emph{upstreamness} and \emph{downstreamness} to quantify the position of each economic sectors (and countries as a whole) with respect to final \textcolor{black}{demand}, and primary factors of production, respectively (see Section \ref{sec:Definitions} for details). 

In a recent paper that has attracted much attention, Antr\`as and Chor \cite{Antras2018} reported empirical observations of a puzzling correlation existing between the upstreamness and downstreamness of several countries\footnote{The upstreamness (or downstreamness) of a country is a weighted average upstreamness (downstreamness) of the economic sectors of the country (see Section \ref{sec:Definitions} for details).} over many years (already noted in \cite{Miller}). More precisely, they used data from the World Input-Output Database (WIOD) for the period 1995-2011, and observed that ``countries that appear to be upstream according to their production-staging distance from final demand (U) are at the same time recorded to be downstream according to their production-staging distance from primary factors (D)'', meaning that ``countries that sell a disproportionate share of their output directly to final consumers (thus appearing to be downstream in GVCs according to U) tend to also feature high value-added over gross output ratios, reflecting a limited amount of intermediate inputs embodied in their production (thus appearing to be upstream in GVCs according to D)''. A scatter plot of upstreamness vs. downstreamness at country level shows an evident linear relation with slope close to $+1$, an effect that persisted in all years of their sample -- and that even intensified between 1995-2011 (see e.g. Figs. 4 and 5 in \cite{Antras2018}). Similar effects are then shown also at the single country-industry level (see e.g. Fig. 10 in \cite{Antras2018}).

Several explanatory factors have been put forward in \cite{Antras2018} to make sense of these puzzling correlations, notably the possible persistence of large trade barriers across countries -- which is however ruled out, as trade costs were found to have fallen off significantly over the period 1995-2011 -- and the growing importance of the service sectors, which typically feature short production chain lengths and little use of intermediate inputs in production. We also mention that the standard definitions of upstreamness and downstreamness have been critically re-assessed, e.g. in \cite{criticaAntras}, and alternative measures put forward there \textcolor{black}{(see also \cite{alternative1} for an alternative proposal based on a dynamical model of supply chains rooted in evolutionary game theory)}. \textcolor{black}{Our analysis lends further support to the warning issued in \cite{criticaAntras} concerning the subtleties involved in the traditional measures of upstreamness and downstreamness when interpreted as simple proxies for a sector's (country's) mere positioning along the GVC.}

Besides looking at empirical data, it is sensible to corroborate the analysis with a complementary approach, namely the use of random models of interconnected economies, which have had a long and fruitful history in econometric studies \cite{Peterson,McNerney2018,McNerneyThesis2012,KopJansen1994,KopJansen1990,Evans1954,Quandt1958,Simonovits1975,West1986,Kogelschatz2007,Kozicka2019,Katz,burford,drake,Phibbs,resurrection,requiem,alive}. The rationale is that whatever empirically observed effect survives randomization of the pairwise interaction between constituents cannot be due to any tailored and specific piece of information carried by the data, but must instead be generic and only due to global and structural constraints. In this spirit, we propose to look at the reported puzzling correlations between upstreamness and downstreamness in Global Value Chains through the prism of (i) a random model of I-O tables, whose entries are drawn independently at random from a given distribution, preserving a few minimal structural constraints (essentially, non-negativity of the entries, and \textcolor{black}{row} sub-stochasticity\footnote{\textcolor{black}{This means that the sum of elements in each row is smaller than or equal to $1$.}} \textcolor{black}{of a closely related matrix, see Section \ref{sec:Definitions} below}), for which the correlation between upstreamness and downstreamness can be tackled analytically, and (ii) a randomized process whereby the columns of an empirical I/O table where such correlations were detected are randomly reshuffled, therefore preserving the row sums of the original matrix. In both cases, we wish to see if randomization of the inter-sectorial dependencies destroys such correlations, as it would be natural to expect if these were due to finely tuned and subtle economic considerations. Contrary to our expectations, though, we find overwhelming evidence that it actually does not. 

\textcolor{black}{Investigating further the economic factors that lead to correlations as hard to destroy as difficult to make intuitive sense of, we find that a crucial role is played by the \textcolor{black}{input-output identities} expressing the fundamental equilibrium rule that ``outputs + final demand'' should perfectly balance ``inputs + value added'' at the sectorial level. \textcolor{black}{We will refer in the following to such accounting identity as \emph{input-output identity}}. Such identities not only
generate a \emph{de facto} denser graph of interactions between sectors, which in turn tends to bring U and D closer to each other, but they also blur the natural interpretation one would expect to assign to these measures (at least in simple GVC topologies), namely as proxies for the positioning of each sector along the production chain -- so that sectors closer to the final demand should have low $U$ \emph{but simultaneously high} $D$, and vice versa for sectors closer to primary factors of production. We will see that this natural intuition may actually fail even in simple examples, a further indication that such measures should collectively be interpreted with much caution.}

The paper is organized as follows. In Section \ref{sec:Definitions} we provide the technical background, including the definition and interpretation of upstreamness and downstreamness, and how these measures are constructed from the I-O table. \textcolor{black}{In Section \ref{sec:curse} we will analyze in detail the simplest case of a linear production chain to show that the \textcolor{black}{input-output identities} (in the form of labor injected in the production process) (i) forces the downstreamness values to deviate from the na\"ive interpretation of positioning index along the chain starting from primary factors of production, and (ii) together with the non-negativity constraint on I/O matrix elements, it 
pushes U and D to be more aligned with each other, effectively preventing these measures from pointing in opposite directions (as it would be quite natural to expect) unless the production chain is truly trivialized -- a set of consequences that we collectively dub ``the curse of the \textcolor{black}{input-output identity}''.} In Section \ref{sec:RandomModel} we first assume that the interaction matrix $A$ between sectors is a random matrix, and then we construct the corresponding Upstreamness and Downstreamness matrices as well as the ``covariance'' and ``slope'' observables that we can monitor numerically and compute analytically in some cases. In Section \ref{sec:results}, we provide our analytical results on a random model with exponential disorder, which shows that the scatter plot between upstreamness and downstreamness of the same sector has necessarily slope $+1$ for any matrix size $N$. These results are tested numerically in Section \ref{sec:NumericalSimulations}, along with numerical tests for other distributions of the entries of $A$, all confirming the same conclusions. In Section \ref{sec:reshuffling} we perform our random reshuffling experiment on empirical I-O matrices, which demonstrates that matrices satisfying the same structural constraints as the original one, but with any real economic information about inter-sectorial relations being wiped out, still display the same strong correlations between upstreamness and downstreamness as the original interaction matrix. Finally, in Section \ref{sec:reshuffling} we offer some critical discussion and concluding remarks.

\section{Definition of Upstreamness and Downstreamness}
\label{sec:Definitions}
\begin{figure}[h]
    \centering
    \includegraphics[width=0.5\textwidth]{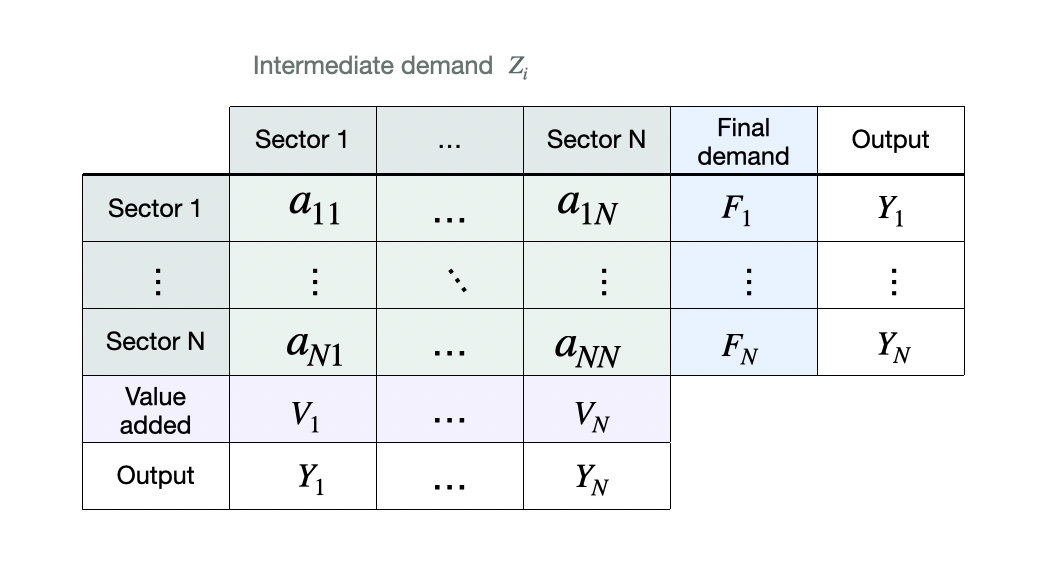}
    \caption{Scheme of the structure of a single-country input-output table \cite{wiotdataset,handbook, suganuma}.}
    \label{fig:ioscheme2}
\end{figure}

Antr\`{a}s and Chor  \cite{antras2012econometrica} considered a closed economy of $N$ industries with no inventories -- for instance, corresponding to a hypothetical single country that does not trade with others. For each industrial sector $i\in\{1,2,\ldots,N\}$ the value of gross output indicated with $Y_i$ equals the sum of its use as a final good ($F_i$) and its use as an intermediate input to other industries ($Z_i$)
\begin{align}
Y_i&= F_i +Z_i = F_i+\sum_{j=1}^N a_{ij}\label{eq:iteration1} \\
&=F_i +\sum_{j=1}^N d_{ij}Y_j \label{eq:iteration}\ .
\end{align}
Here, \textcolor{black}{$a_{ij}$ is the total value in monetary units (e.g. US dollars) of $i$'s output used to produce $j$'s output, while} $d_{ij}$ is the amount \textcolor{black}{of monetary units} of sector $i$'s output needed to produce one \textcolor{black}{monetary unit}'s worth of sector $j$'s output (see schematic structure of a I-O matrix for a single country in Fig. \ref{fig:ioscheme2}). 

Iterating the identity in \eqref{eq:iteration}, one obtains an infinite sequence of terms reflecting the use of  sector $i$'s output at different level in the value chain
\begin{equation}
  Y_i = F_i + \sum_{j=1}^N d_{ij}F_j + \sum_{j=1}^N \sum_{k=1}^N d_{ik}d_{kj}F_j +\dots\ . \label{eq:recursion}
\end{equation}
Using the matrix geometric series $\sum_{k\geq 0}D^k = [\mathds{1}_N-D]^{-1} $, we can eventually rewrite \eqref{eq:recursion} as
\begin{equation}
    {\bm Y} = [\mathds{1}_N-D]^{-1}{\bm F}\ ,\label{eq:recursionmatrix}
\end{equation}
where $\mathds{1}_N$ is the $N\times N$ identity matrix, $D=(d_{ij})$ is the matrix of \textcolor{black}{monetary units} values, and $\bm{F}$ the column vector of final demands.

 Antr\`{a}s and Chor \cite{antras2012econometrica} therefore proposed the following measure of upstreamness of the $i$-th industry, by multiplying each of the terms in \eqref{eq:recursion} by their distance from final use, and dividing by $Y_i$
\begin{align}
\nonumber U_{1i} &= 1 \cdot \frac{F_i}{Y_i} + 2 \cdot\frac{\sum_{j=1}^N d_{ij}F_j}{Y_i} \\
&+ 3 \cdot \frac{\sum_{j,k=1}^N d_{ik}d_{kj}F_j}{Y_i} + \dots = \frac{([\mathds{1}_N - D]^{-2}{\bm F})_i}{Y_i} \ ,
\label{eq:up0}\end{align}
where $(\cdot)_i$ indicates the $i$-th component of the vector.

Inserting \eqref{eq:recursionmatrix} into \eqref{eq:up0}, we can rewrite the upstreamness vector as
\begin{equation}
  {\bm U_1} =  [\mathds{1}_N-A_U]^{-1}{\bm 1} \ , \label{eq:up1}
\end{equation}
where
\begin{equation}
A_U=  Y^{-1}A = \begin{pmatrix} \frac{a_{11}}{Y_1} & \cdots & \frac{a_{1N}}{Y_1} \\
\vdots& \ddots & \vdots  \\
\frac{a_{N1}}{Y_N} & \cdots & \frac{a_{NN}}{Y_N}  \end{pmatrix} \label{eq:AU}
\end{equation}
and $Y=\mathrm{diag}(Y_1,\ldots,Y_N)$. The matrix $A_U$ has therefore non-negative elements, and is \emph{row-substochastic} ($\sum_j (A_U)_{ij}\leq 1$ for all sectors $j$) because $(A_U)_{ij}=d_{ij}Y_j/Y_i$ is the share of sector $i$'s total output that is purchased by industry $j$.

The upstreamness is defined in such a way that terms of the sum that are further upstream in the value chain have larger weight. By construction $U_{1i}\geq 1$ and is precisely equal to $1$ if no output of industry $i$ is used as input to other industries, that is the output of industry $i$ is only used to satisfy the final demand.

Antr\`{a}s et al. \cite{AntrasFally2012} later established an equivalence between their upstreamness measure and a measure -- defined in a recursive fashion -- of the ``distance" of an industry from the final demand proposed independently by Fally \cite{Fally2012}. Fally's upstreamness $U_2$ is defined as follows:
\begin{equation}
    U_{2i} = 1 + \sum_{j=1}^N\frac{d_{ij}Y_j}{Y_i}U_{2j} \ .
    \label{eq:fally1}
\end{equation}
The idea is that $\bm U_2$ aggregates information on the extent to which a sector in a given country produces goods that are sold directly to final consumers or that are sold to other sectors that themselves sell largely to final consumers.  Sectors selling a large share of their output to relatively upstream industries should be therefore relatively upstream themselves.
Using the fact that $d_{ij}Y_j = a_{ij}$ we have again that
\begin{equation}
  {{\bm U}_2}= [\mathds{1}_N-A_U]^{-1}{{\bm 1}}\ ,
        \label{eq:fally2}
\end{equation}
where $A_U$ is defined in \eqref{eq:AU}. 
In \cite{Antras2018} an application of those measures for the analysis of empirical data on global value chains is presented.

On the input-side, there is an  \textcolor{black}{input-output} accounting identity that sector $i$’s total input $Y_i$ be equal to the value of its primary inputs (value added) $V_i$ plus its intermediate input purchases from all other sectors: 
\begin{equation}
   Y_i= V_i +Z_i = V_i +\sum_{j=1}^N a_{ji}= V_i +\sum_{j=1}^N d_{ji}Y_j \ ,\label{accid2}
\end{equation}
or in vector/matrix form
\begin{equation}
  {\bm Y}= [\mathds{1}_N-D^T]^{-1}{\bm V}\ . \label{eq:inputchain}
\end{equation}
Similarly to \cite{antras2012econometrica} (see \eqref{eq:up0}), Miller and Temurshoev \cite{Miller} introduced the so-called \emph{downstreamness} measuring the average distance between suppliers of primary inputs and sectors as input purchaser along the input demand supply chain as follows
\begin{align}
\nonumber   D_{1i} &= 1 \cdot \frac{V_i}{Y_i} + 2\cdot\frac{\sum_{j=1}^N V_j d_{ji}}{Y_i} +\\
&+3\cdot \frac{\sum_{j,k=1}^N V_j d_{jk}d_{ki} }{Y_i} + \dots
=   \frac{([\mathds{1}_N - D^T]^{-2}{\bm V})_i}{Y_i} \ .
\end{align}
As before, using \eqref{eq:inputchain}, we obtain
\begin{equation}
    {\bm D}_1= [\mathds{1}_N -A_D]^{-1}{\bm 1} \label{eq:D1}
\end{equation}
with \begin{equation}
    \ \ {\rm with} \ A_D= (A Y^{-1})^T = \begin{pmatrix} \frac{a_{11}}{Y_1} & \cdots & \frac{a_{N1}}{Y_1} \\
\vdots& \ddots & \vdots  \\
\frac{a_{1N}}{Y_N} & \cdots & \frac{a_{NN}}{Y_N}  \end{pmatrix}\ . \label{eq:AD}
\end{equation}
Also the matrix $A_D$ has therefore non-negative elements, and is \emph{row-substochastic} ($\sum_j (A_D)_{ij}\leq 1$ for all sectors $j$). It is worth noting that by construction the matrices $A_U$ and $A_D$ share the diagonal elements $a_{ii}/Y_i$.

Finally, as in the upstreamness case, also for the downstreamness, Fally \cite{Fally2012} introduced an analogous iterative definition of the form

\begin{equation}
    D_{2i} = 1 + \sum_{j=1}^N d_{ji}D_{2j} \ ,  \label{eq:D2}
\end{equation}
which can be again mapped with simple manipulations into Eq. \eqref{eq:D1} using $Y_i d_{ji}=a_{ji}$.

The I-O Table in Fig. \ref{fig:ioscheme2} can be modified in a conceptually simple way to account for inter-country trade by accommodating different inter-sectorial blocks (one for each country) -- see scheme in Fig. 1 of \cite{Antras2018}. The upstreamness (or downstreamness) of a country is then a suitably averaged (aggregate) version of the upstreamness (or downstreamness) of all industrial sectors of that country. In principle, there are two different ways to perform this aggregation. First, one could take the ``giant'' I-O table and collapse its entries at the country-by-country level by computing the total purchases of intermediate inputs by country $j$ from country $i$ -- and then compute the upstreamness and the downstreamness on the collapsed (aggregate) table. Or, one could keep working with the giant table, compute the upstreamness and the downstreamness of industrial sectors within a country, and then perform a suitable average of those at country level. In \cite{Antras2018}, the two approaches were found to deliver extremely highly correlated country-level indices of GVC positioning.

\section{\textcolor{black}{The curse of the \textcolor{black}{input-output identity} }}\label{sec:curse}

\begin{figure}[h]
    \centering
    \includegraphics[width=0.43\textwidth]{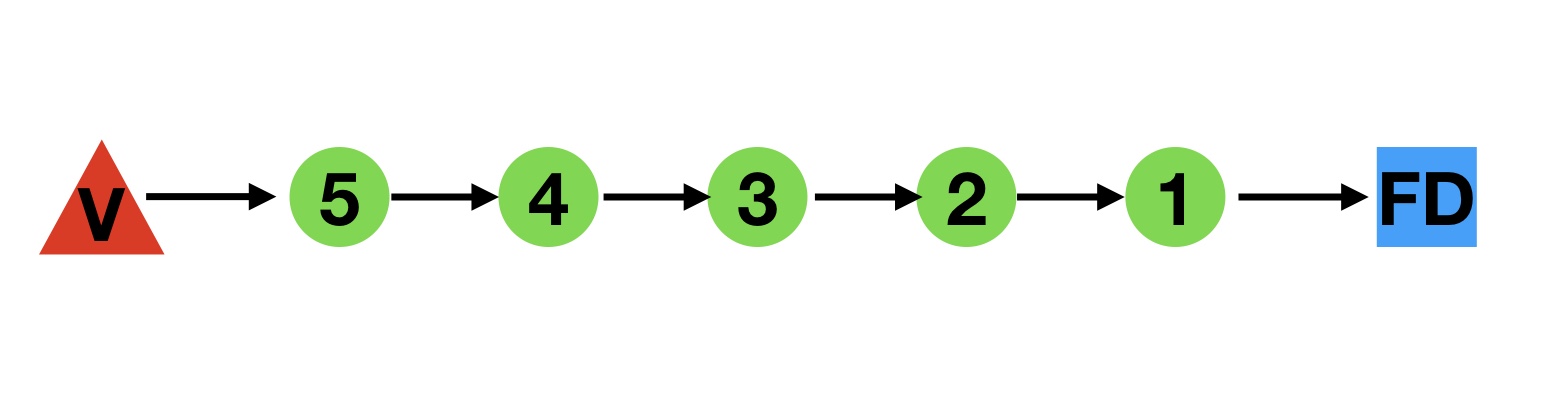}
    \caption{\textcolor{black}{Schematic representation of a production chain of five sectors, where each sector receives goods from a single other sector (or primary factors of production, in the case of $5$), and transfer it to another one (or to the consumer to meet the Final Demand, in the case of $1$). }}
    \label{fig:chain}
\end{figure}

\textcolor{black}{In this Section, we illustrate a few subtleties with the interpretation of upstreamness and downstreamness defined earlier as simple indicators of a sector's positioning along the production process by considering the simplest example of a production (linear) chain, where sectors are arranged as in Fig. \ref{fig:chain}.}

\textcolor{black}{Each sector receives goods as input from a single other sector, while sector $1$ sells directly to final consumers, and sector $5$ handles primary factors of production directly.}

\textcolor{black}{In such a simple geometry, we would expect to be able to read off the upstreamness ($\bm U_1 = (1,2,3,4,5)^T$) and downstreamness  ($\bm D_1 = (5,4,3,2,1)^T$) values of each sector by simply identifying their positioning along the production chain with respect to final consumers and primary factors of production, respectively. However, we will see shortly that this natural expectation is violated even in this simple geometry due to the \textcolor{black}{input-output identity} (i.e. the need to take into account labor injected in the production chain at every step). }

\textcolor{black}{The I/O matrix $A$ that corresponds to the chain in Fig. \ref{fig:chain} reads 
\begin{equation}
    A=\begin{pmatrix}
    0 & 0 & 0 & 0 & 0\\
    a_{21} & 0 & 0 & 0 & 0\\
    0 & a_{32} & 0 & 0 & 0\\
    0 & 0 & a_{43} & 0 & 0\\
    0 & 0 & 0 & a_{54} & 0\\
    \end{pmatrix}\ ,
\end{equation}
with corresponding final demand vector $\bm F = (F_1,0,0,0,0)^T$, with $F_1\neq 0$, to reflect the fact that the sector $1$ does not transfer output to any other sectors, but only to the final consumer, while all other sectors do not interact directly with the final consumer. As a consequence, the matrices $A_U$ and $A_D$ defined in Eqs. \eqref{eq:AU} and \eqref{eq:AD} read
\begin{equation}
    A_U=\begin{pmatrix}
    0 & 0 & 0 & 0 & 0\\
    1 & 0 & 0 & 0 & 0\\
    0 & 1 & 0 & 0 & 0\\
    0 & 0 & 1 & 0 & 0\\
    0 & 0 & 0 & 1 & 0\\
    \end{pmatrix}
\end{equation}
and 
\begin{equation}
   A_D=\begin{pmatrix}
    0 & a_{21}/F_1 & 0 & 0 & 0\\
    0 & 0 & a_{32}/a_{21} & 0 & 0\\
    0 & 0 & 0 & a_{43}/a_{32} & 0\\
    0 & 0 & 0 & 0 & a_{54}/a_{43}\\
    0 & 0 & 0 & 0 & 0\\
    \end{pmatrix}\ .
\end{equation}
Interestingly, by imposing the chain topology \emph{from the viewpoint of interaction with final consumer} (i.e. fixing the vector of final demands while leaving the value added vector $\bm V$ unconstrained), we trivialize the matrix $A_U$ (killing any parametric dependence on the actual ``weights'' of the goods transferred along the chain). }

\textcolor{black}{We can now compute the upstreamness and downstreamness vectors $\bm U_1$ and $\bm D_1$ from the definitions in Eqs. \eqref{eq:up1} and \eqref{eq:D1} obtaining
\begin{align}
  \bm U_1 &= (1,2,3,4,5)^T\\
  \bm D_1 &=
  \begin{pmatrix}
  1+\frac{a_{21}}{F_1}+\frac{a_{32}}{F_1}+\frac{a_{43}}{F_1}+\frac{a_{54}}{F_1}\\
  1+\frac{a_{32}}{a_{21}}+\frac{a_{43}}{a_{21}}+\frac{a_{54}}{a_{21}}\\
  1+\frac{a_{43}}{a_{32}}+\frac{a_{54}}{a_{32}}\\
  1+\frac{a_{54}}{a_{43}}\\
  1
  \end{pmatrix}\ .\label{D1chain}
\end{align}
Perhaps unexpectedly, while the upstreamness vector precisely returns the positioning of each vector along the chain with respect to final consumers, the downstreamness vector has a more complicated and less intelligible structure, which does not match our na\"ive expectation $\bm D_1=(5,4,3,2,1)^T$. Incidentally, this point was mentioned in passing also in the paper \cite{criticaAntras}, where the authors stated: ``Curiously, sector rankings by ``upstreamness'' and ``downstreamness'' measures do not coincide with each other. This implies certain inconsistency in the way that these measures are defined.'' }

\textcolor{black}{It is interesting to further stress the following:
\begin{enumerate}
    \item We have so far ignored the \textcolor{black}{input-output identities} \eqref{eq:iteration1} and \eqref{accid2} that further imply
    \begin{equation}
        V_i +\sum_{j=1}^N a_{ji}=F_i +\sum_{j=1}^N a_{ij}
    \end{equation}
    for every sector $i$. Specializing to our linear chain, we obtain the following set of identities
    \begin{align}
 \nonumber       V_1+a_{21} &=F_1\\
\nonumber        V_2+a_{32} &=a_{21}\\
 \nonumber       V_3+a_{43}&=a_{32}\\
  \nonumber      V_4+a_{54}&=a_{43}\\
       V_5 &=a_{54}\ ,\label{accountingcurse}
    \end{align}
    which can be visually represented as additional directed links (see Fig. \ref{fig:chain2}).
    \begin{figure}[h]
    \centering
    \includegraphics[width=0.43\textwidth]{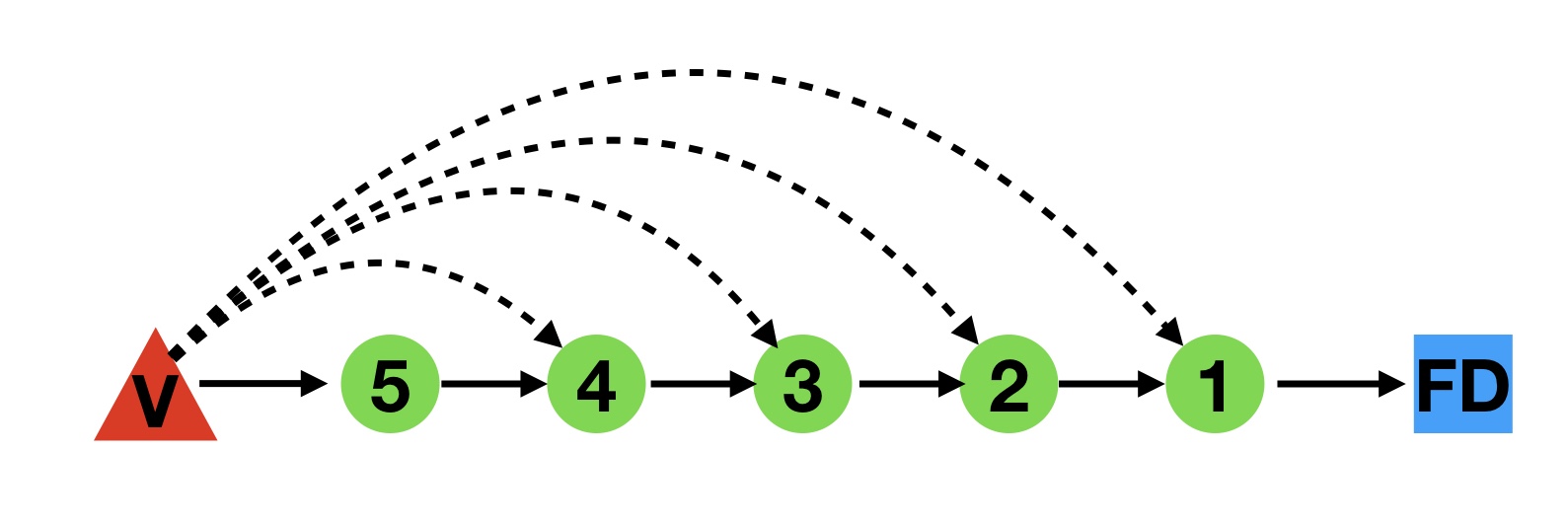}
    \caption{\textcolor{black}{Schematic representation of the same production chain of five sectors as in Fig. \ref{fig:chain}, this time including the value added at each node, for instance in the form of labor utilized to process goods. }}
    \label{fig:chain2}
\end{figure}
\item The ``natural'' downstreamness vector $\bm D_1=(5,4,3,2,1)^T$ (representing simply the positioning of sectors along the chain with respect to primary factors of production) could be obtained from \eqref{D1chain} only if $a_{21}=a_{32}=a_{43}=a_{54}=F_1$, which would in turn imply $V_1=V_2=V_3=V_4=0$ and $V_5\neq 0$ from the \textcolor{black}{input-output identities} above. This is a highly degenerate case where sector $5$ handles primary factors of production and then transfer them seamlessly along the chain, with all other sectors playing just the role of ``paper pushers'' until the unprocessed goods reach the final consumer. In all other cases where sectors actually operate injecting labor on the incoming goods, the downstreamness level of each sector may get distorted compared to their basic positioning, to the point that even the \emph{relative} ordering between sectors (e.g. sector $1$ should be more downstream than sector $3$) may be subverted. Take for instance $a_{21}=1.5$, $a_{32}=\alpha a_{21}$, $a_{43}=\alpha^2 a_{21}$, $a_{54}=\alpha^3 a_{21}$, with $\alpha=1/3$ and $F_1=100$. This choice leads to the downstreamness of sector $1$ to be \emph{lower} than that of sectors $2,3,4$, which seems paradoxical at first sight. However, this happens because the \textcolor{black}{input-output identities} \eqref{accountingcurse} force $V_1$ to be $\sim 100$ times larger than $V_2$, and $\sim 1000$ times larger than $V_3$ and $V_4$: a large external demand, coupled with a relatively low volume of incoming goods to process, forces a strong demand for labor at sector $1$, which in turn considerably lowers its expected downstreamness value given its position along the chain (farthest sector from primary factors of production). 
\end{enumerate}}
\textcolor{black}{In summary, this simple example already shows that $U$ and $D$ do not simply reflect the positioning of a sector (even along a linear production chain), since the \textcolor{black}{input-output identities} force an effectively denser network geometry, with extra links corresponding to value added injected at each step (compare Figs. \ref{fig:chain} and \ref{fig:chain2}).} 

\textcolor{black}{Moreover, we can study the $U$-$D$ Pearson correlation coefficient $\rho\in [-1,1]$. At least in this simple geometry, we should expect a (maximally) \emph{negative} correlation (close to $-1$), irrespective of the actual production parameters -- as sectors closer to final demand appear to be automatically farther from primary factors of production. Surprisingly, we find that the economic parameters can be chosen to produce \emph{any} value of $\rho$ (including \emph{positive} values!): a sector can therefore be simultaneously quite upstream with respect to final demand, and quite downstream with respect to primary factors of production, unless there is no value added along the chain (the ``paper pushers'' scenario described above), in which the two measures are perfectly anticorrelated (see below).}

\begin{figure}[h]
    \centering
    \includegraphics[width=0.47\textwidth]{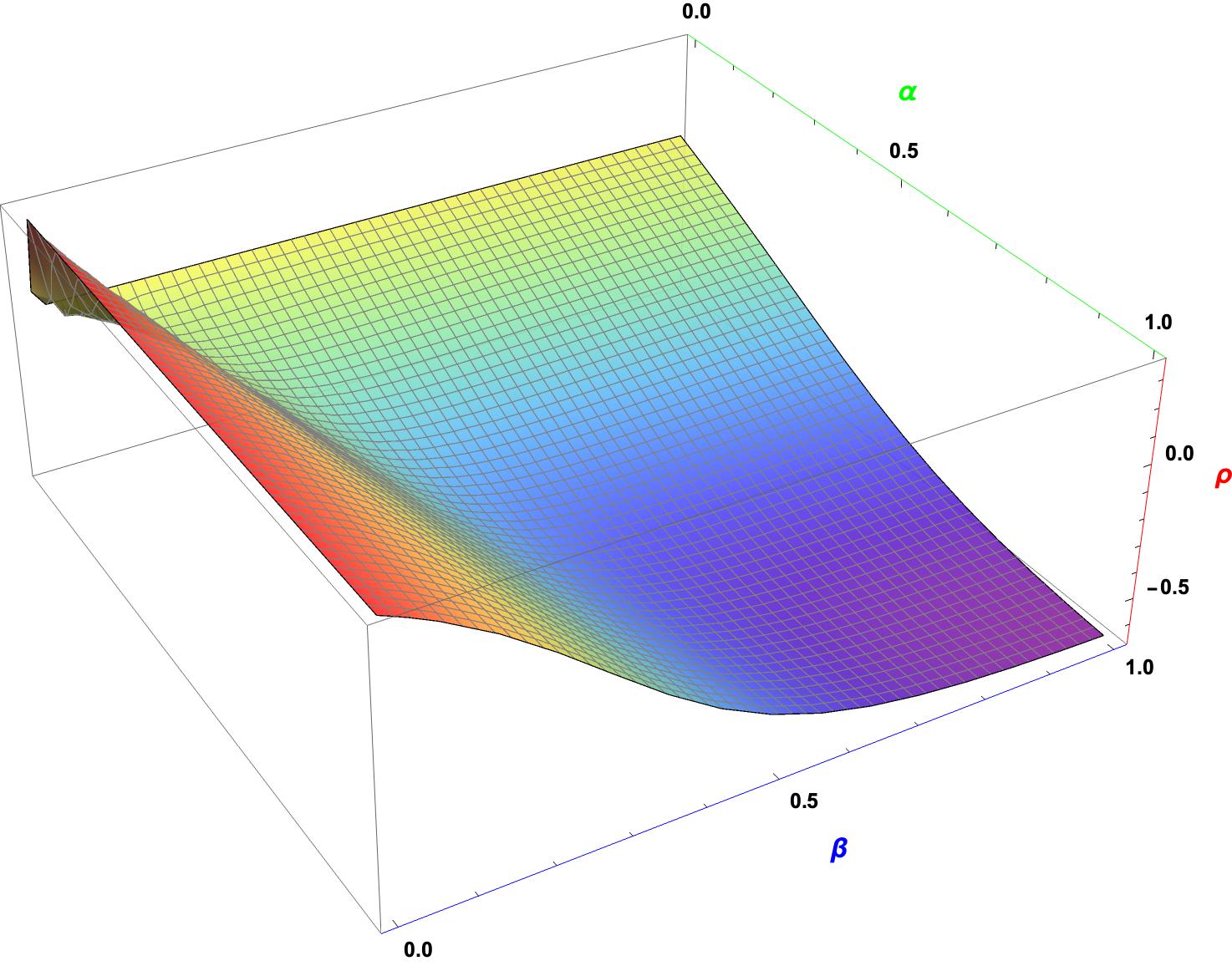}
    \caption{\textcolor{black}{U-D Pearson correlation coefficient $\rho$ for the simplified scenario described in the main text, with $t=1/3$, as a function of $\alpha$ and $\beta$.}}
    \label{fig:rhopearson}
\end{figure}

\textcolor{black}{To make the calculation more transparent, we put ourselves in a simplified scenario and take $a_{21}$ as reference point, setting $a_{32}=a_{54}=\alpha a_{21}$, $a_{43}=\beta a_{21}$, and $a_{21}/F_1=t$. The \textcolor{black}{input-output identities} \eqref{accountingcurse} force the elementary constraints $0\leq \alpha,\beta,t\leq 1$. The Pearson correlation coefficient $\rho$ has a lengthy but fully explicit expression as a function of $\alpha,\beta,t$, given by $\rho=\Omega_1/\Omega_2$, with
\begin{equation}
    \Omega_1= \alpha -2 \alpha  \beta  (2 t+1)-\beta  (\beta +2 (\beta +1) t)
\end{equation}
and}
\begin{strip}
\begin{dgroup*}
\begin{dmath*}
\textcolor{black}{\Omega_2=2 \beta  \left[\alpha ^2 \left(\frac{2-2 \beta }{\beta ^2}+8\right)+\frac{2 \beta ^2}{\alpha ^2}+\alpha  \left(8 \beta -\frac{1}{\beta }-3\right)\right.}\\
\textcolor{black}{\left.-\frac{(\beta -4) \beta }{\alpha }+\beta  (2 \beta -3)+2 t^2 (2 \alpha +\beta +1)^2-\frac{t (2 \alpha +\beta +1) \left(\alpha ^2 (2 \beta +1)+\alpha  \beta  (\beta +1)+\beta ^2\right)}{\alpha  \beta }+1\right]^{1/2}\ ,}
\end{dmath*}
\end{dgroup*}
\end{strip}
\textcolor{black}{which can be plotted and analyzed in detail (see Fig. \ref{fig:rhopearson} for $t=1/3$).}

\textcolor{black}{First, we observe that the limit $t,\alpha,\beta\to 1$ corresponds to sending all added values $V_i$ to zero, reproducing the ``paper pusher'' scenario described above: in this case, the Pearson correlation coefficient $\rho$ correctly tends to $-1$, signaling that in this case U and D are maximally anti-correlated (the closest sector to final demand must be the farthest sector from primary factors of production).}

\begin{figure}[h]
    \centering
    \includegraphics[width=0.44\textwidth]{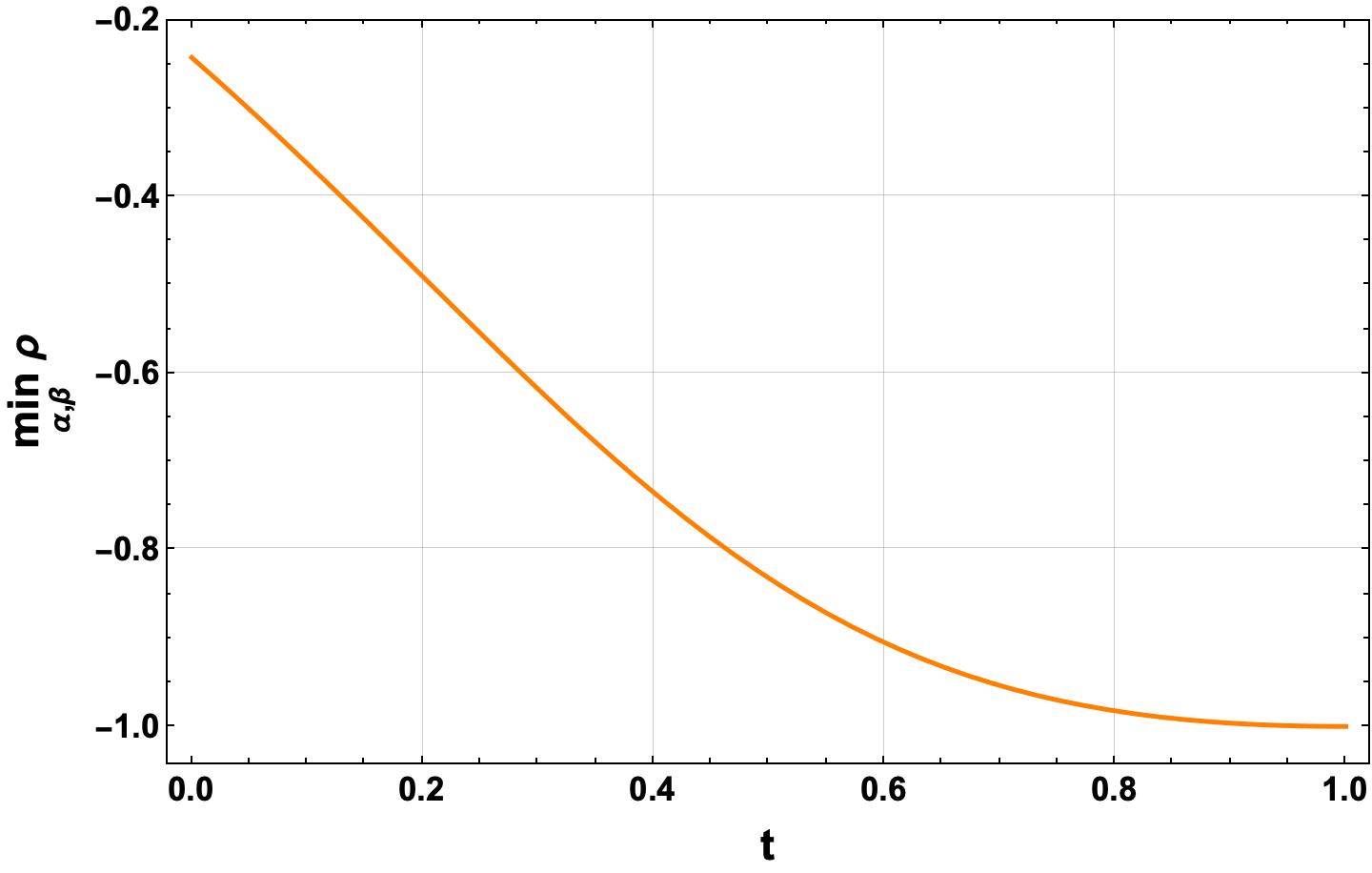}
    \caption{\textcolor{black}{Minimal value of the $U$-$D$ Pearson correlation coefficient as a function of $t$. For $t$ strictly smaller than $1$, $\rho$ is lower bounded away from $-1$.}}
    \label{fig:minrho}
\end{figure}

\textcolor{black}{However, a more striking observation is that, for $t$ strictly smaller than $1$, the minimum value of $\rho$ (while still negative), is bounded away from $-1$ (see Fig. \ref{fig:minrho}), which means that having just $V_1$ strictly larger than zero is already enough to push U and D closer to each other. Moreover, the maximum value of $\rho=1/2\sqrt{2}\approx 0.3535...$ -- obtained for vanishing $\beta$, irrespective of $\alpha,t$ -- is \emph{positive} (and all intermediate values are possible). So, even in a perfectly linear chain, where the positioning of a sector according to final demand should be perfectly specular to its positioning according to primary factors of production, it is eminently possible to have U and D weakly anti-correlated, uncorrelated, or even positively correlated instead, for instance provided one of the links is sufficiently weak that the value added at the corresponding node must ``step up'' to improve the flow along the chain.}

\textcolor{black}{In summary, the simple ``chain'' geometry already shows that the natural interpretation of U and D as measures of each sector's positioning along the production process must be handled with care, and that a positive correlation between the two (or at least a much weaker anti-correlation than one might na\"ively expect) may naturally arise because of the effectively denser structure of the production chain once value added (e.g. in the form of labor) is included in the picture due to the \textcolor{black}{input-output identities}  \eqref{accountingcurse}. In this context, it is worth highlighting that this ``weakened anti-correlation'' was precisely at the heart of a different proposal to measure the relative positioning of a sector within GVC, put forward in \cite{criticaAntras}. Citing \cite{criticaAntras} again: ``We propose a new production position measure as the relative distance of a particular
production stage (country-sector) \emph{to the both ends}\footnote{\textcolor{black}{Italics added for emphasis.}} of a value chain. Using our definitions, the sector ranking by upstreamness and downstreamness would be exactly \emph{inversely related}. This
removes one inconsistency with the existing measures in the literature. ''}

\textcolor{black}{In the next sections, we will further investigate these correlations in the framework of randomized GVC models.}

\section{The random model}
\label{sec:RandomModel}

Our randomized model is based on the closed-economy paradigm described in \textcolor{black}{Section \ref{sec:Definitions}}, and it assumes that the $N\times N$ matrices $A_U$ and $A_D$ (defined in Eqs. \eqref{eq:AU} and \eqref{eq:AD}, respectively) are generated from a random interaction matrix $A$ between sectors, i.e. without any structural information about the underlying dynamics of goods and prices apart from the constraint that their entries be non-negative, and that the matrices be row-substochastic. See subsection \ref{sec: modeldef} for the precise definition of the random model.

\subsection{Covariance and slope}
Assuming therefore that the underlying model for the interaction matrix $A$ is random, the covariance between the upstreamness $(\bm U_1)_i$ and downstreamness $(\bm D_1)_i$ (defined in Eqs. \eqref{eq:up1} and \eqref{eq:D1}, respectively) of the same $i$-th sector is
\begin{align}
\nonumber    \mathrm{Cov} &((\bm U_1)_i,(\bm D_1)_i)=\\
    &=\mathbb{E}\left[(\bm U_1)_i (\bm D_1)_i\right]-\mathbb{E}[(\bm U_1)_i]\mathbb{E}[(\bm D_1)_i]\ ,\label{eq:full_covariance}
\end{align}
where the expectation $\mathbb{E}[\cdot]$ is taken w.r.t. the joint probability density function (pdf) of the entries of the matrix $A$ (from which $A_U$ and $A_D$ are constructed). Since the upstreamness and downstreamness are defined in terms of a complicated matrix inversion, computing the covariance in Eq. \eqref{eq:full_covariance} is a non-trivial task even for very simple joint pdfs of the entries of $A$.

However, we can take advantage of the results in Ref. \cite{Bartolucci,Bartolucci2}, which demonstrated that the ``true'' upstreamness and downstreamness (as defined in Eqs. \eqref{eq:up1} and \eqref{eq:D1}, respectively) \textcolor{black}{for sufficiently dense matrices} are individually correlated with simpler rank-1 estimators
\begin{align}
\tilde{U}_i &= 1+\frac{r_i}{1-(1/N)\sum_j r_j }\label{eq:tildeUU} \\ 
\tilde{D}_i &= 1+\frac{r_i'}{1-(1/N)\sum_j r_j' }\ ,\label{eq:tildeDD}
\end{align}
where $r_i=\sum_j (A_U)_{ij}$ are the row sums of $A_U$, and $r_i'=\sum_j (A_D)_{ij}$ are the row sums of $A_D$. 

It is therefore sufficient to compute the covariance between the simpler estimators
\begin{equation}
    \mathrm{Cov}(\tilde U_i,\tilde D_i)=\mathbb{E}[\tilde U_i\tilde D_i]-\mathbb{E}[\tilde U_i]\mathbb{E}[\tilde D_i]\label{eq:covariancetilde}
\end{equation}
to draw meaningful conclusions about the covariance between upstreamness and downstreamness as originally defined. 

Noting that the quantities $(1/N)\sum_j r_j$ and $(1/N)\sum_j r_j'$ quickly converge to their non-fluctuating averages $\mathbb{E}[r]$ and $\mathbb{E}[r']$ by virtue of the Law of Large Numbers (LLN), we make the further simplifying move to replace these quantities with their non-fluctuating averages directly in the calculation of the covariance Eq. \eqref{eq:covariancetilde}\footnote{More precisely, we make the approximation $\mathbb{E}\left[\frac{r_i}{1-(1/N)\sum_j r_j }\right]\approx \frac{\mathbb{E}[r]}{1-\mathbb{E}[r]}$, and similarly for $r'$.}. Therefore, our covariance of interest reduces to the following object
\begin{equation}
    \mathcal{C}_N=\frac{\mathbb{E}[rr']-\mathbb{E}[r]\mathbb{E}[r']}{(1-\mathbb{E}[r])(1-\mathbb{E}[r'])}\ ,\label{eq:simplifiedcov}
\end{equation}
where we omitted the $i$-dependence (as every sector is statistically equivalent to any other in our random models). Therefore, $r$ and $r'$ can be viewed as the sum of, say, the first row of $A_U$ and $A_D$, respectively.

\begin{figure}[h]
    \centering
    \fbox{\includegraphics[scale = 0.20]{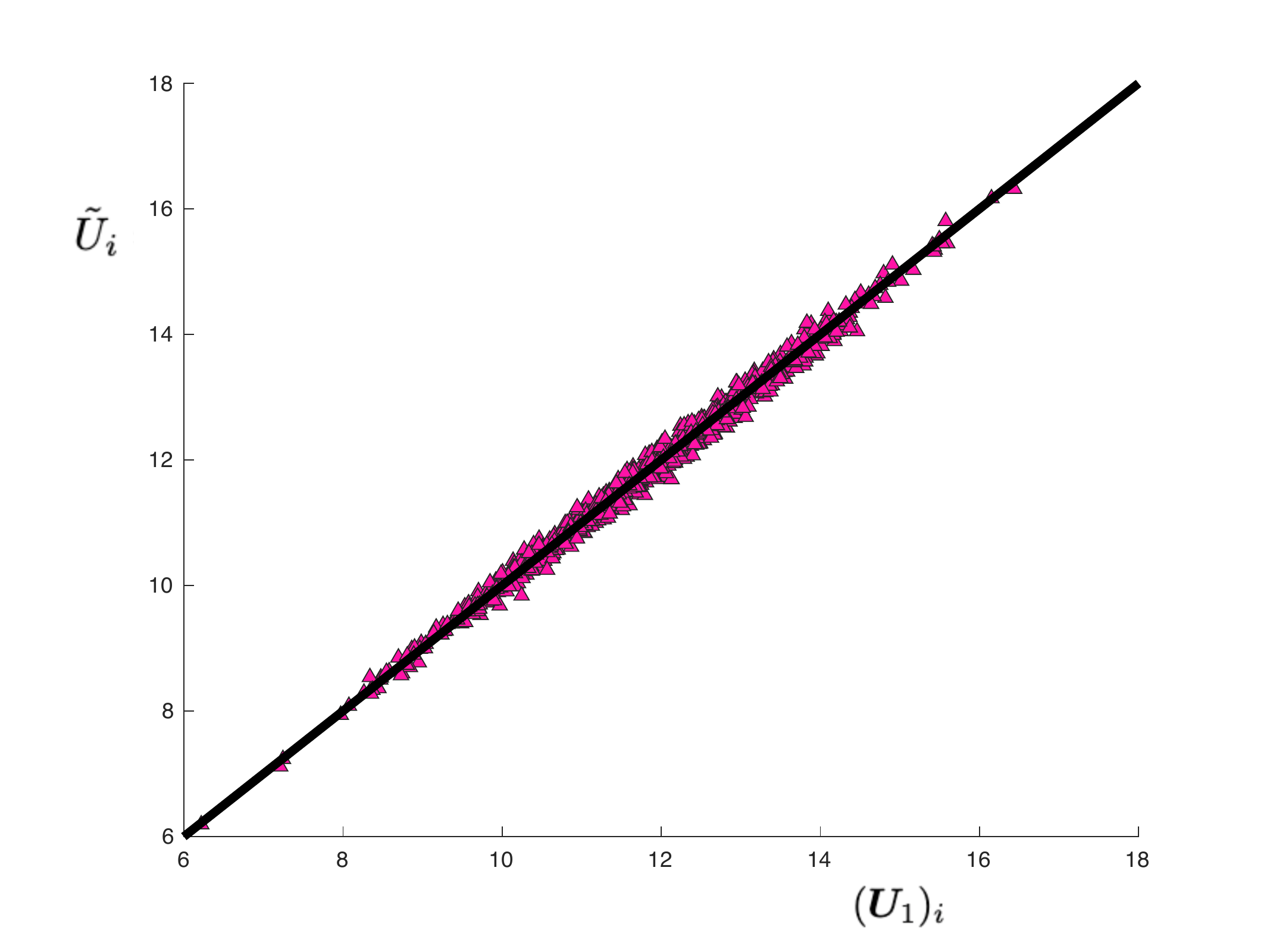}}
    \caption{Scatter plot of the approximate upstreamness (Eq. \eqref{eq:tildeUU}) vs. the ``true'' upstreamness $(\bm U_1)_i$ for our random model with exponential disorder. The parameters used are $\mu=1$, $\mu_F=0.1$, $N=100$, $i=7$. There are $1000$ pairs of points in the figure, each obtained from a different instance of the random matrix $A$ with exponential disorder.}
    \label{fig:scatterU}
\end{figure}

\begin{figure}[h]
    \centering
    \fbox{\includegraphics[scale = 0.20]{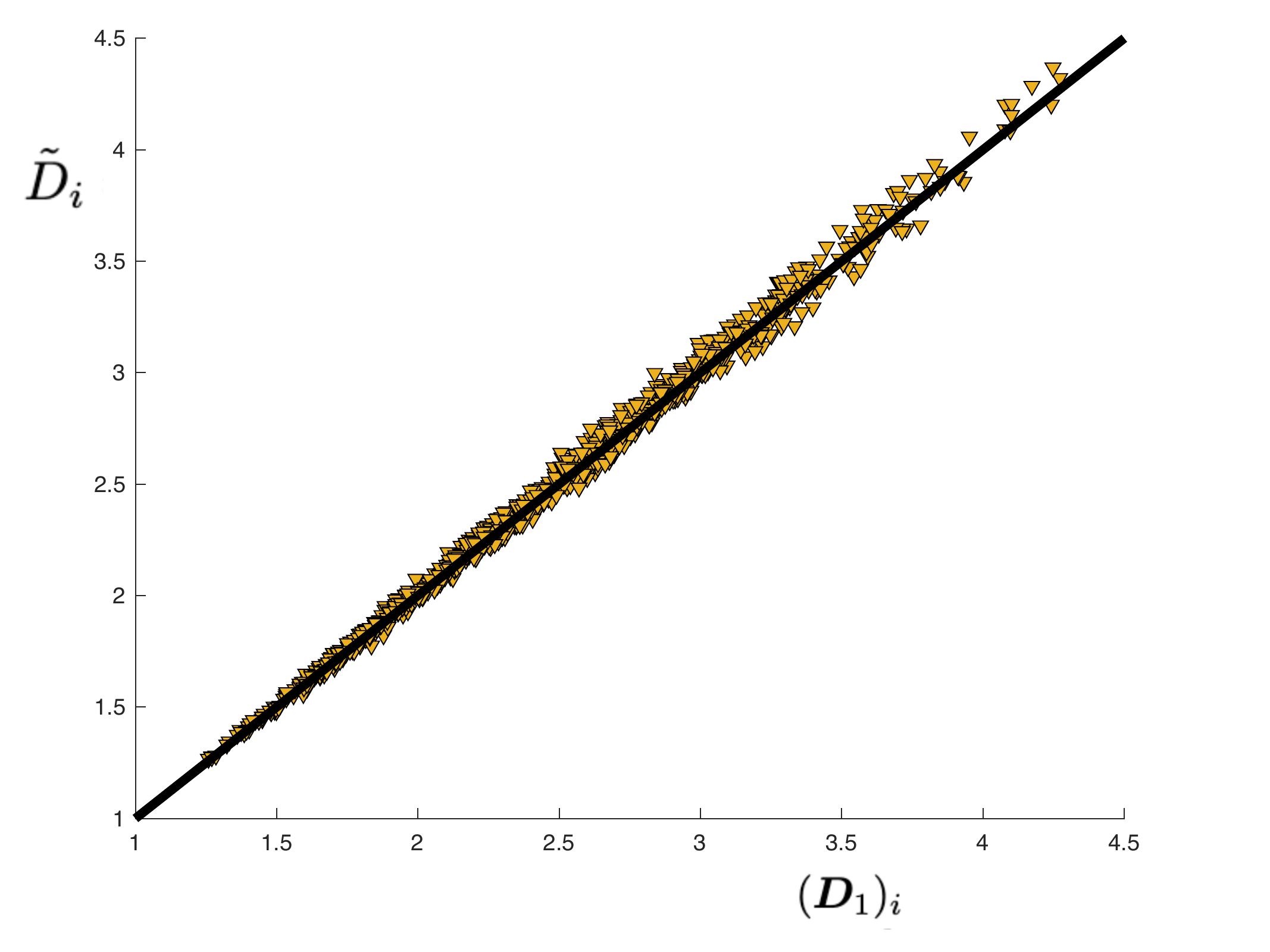}}
     \caption{Scatter plot of the approximate downstreamness (Eq. \eqref{eq:tildeDD}) vs. the ``true'' downstreamness $(\bm D_1)_i$ for our random model with exponential disorder. The parameters used are $\mu=1$, $\mu_F=0.01$, $N=100$, $i=7$. There are $1000$ pairs of points in the figure, each obtained from a different instance of the random matrix $A$ with exponential disorder.}
    \label{fig:scatterD}
\end{figure}

We check with numerical simulations in Figs. \ref{fig:scatterU} and \ref{fig:scatterD} that indeed our conclusions are not affected by the fact that we considered simpler estimators \textit{in lieu} of the original observables, as the former are perfectly correlated with the latter.

The slope $S$ of the scatter plot between $\tilde D_i$ and $\tilde U_i$ is easily determined from Eq. \eqref{eq:covariancetilde} by assuming first that there be a linear relation between the two, $\tilde D_i = S \tilde U_i$, and substituting in the expression for the covariance Eq. \eqref{eq:covariancetilde} we get
\begin{align}
    \mathrm{Cov}(\tilde U_i,\tilde D_i)=
    S\left\{\mathbb{E}[ \tilde{U}_i^2]-\mathbb{E}[ \tilde{U}_i]^2\right\}\ ,
\end{align}
from which we deduce
\begin{equation}
    S = \frac{\mathrm{Cov}(\tilde U_i,\tilde D_i)}{\mathrm{Var}[\tilde{U}_i]}\ ,
\end{equation}
where $\mathrm{Var}[\tilde{U}_i]=\mathbb{E}[ \tilde{U}_i^2]-\mathbb{E}[ \tilde{U}_i]^2$ is the variance of the approximate upstreamness.

Making again the further approximation that $(1/N)\sum_j r_j$ is replaced with its non-fluctuating average $\mathbb{E}[r]$ by virtue of the \textcolor{black}{LLN}, and after simple algebra from \textcolor{black}{Eq. \eqref{eq:covariancetilde}}, we have that the slope $S$ can be approximated by
\begin{equation}
S=\frac{\mathcal{C}_N\left(1-\mathbb{E}[r]\right)^2}{\mathbb{E}[r^2]-\mathbb{E}[r]^2}\ .\label{eq:approximateslope}
\end{equation}

\begin{figure}[h]
    \centering
    \fbox{\includegraphics[scale = 0.20]{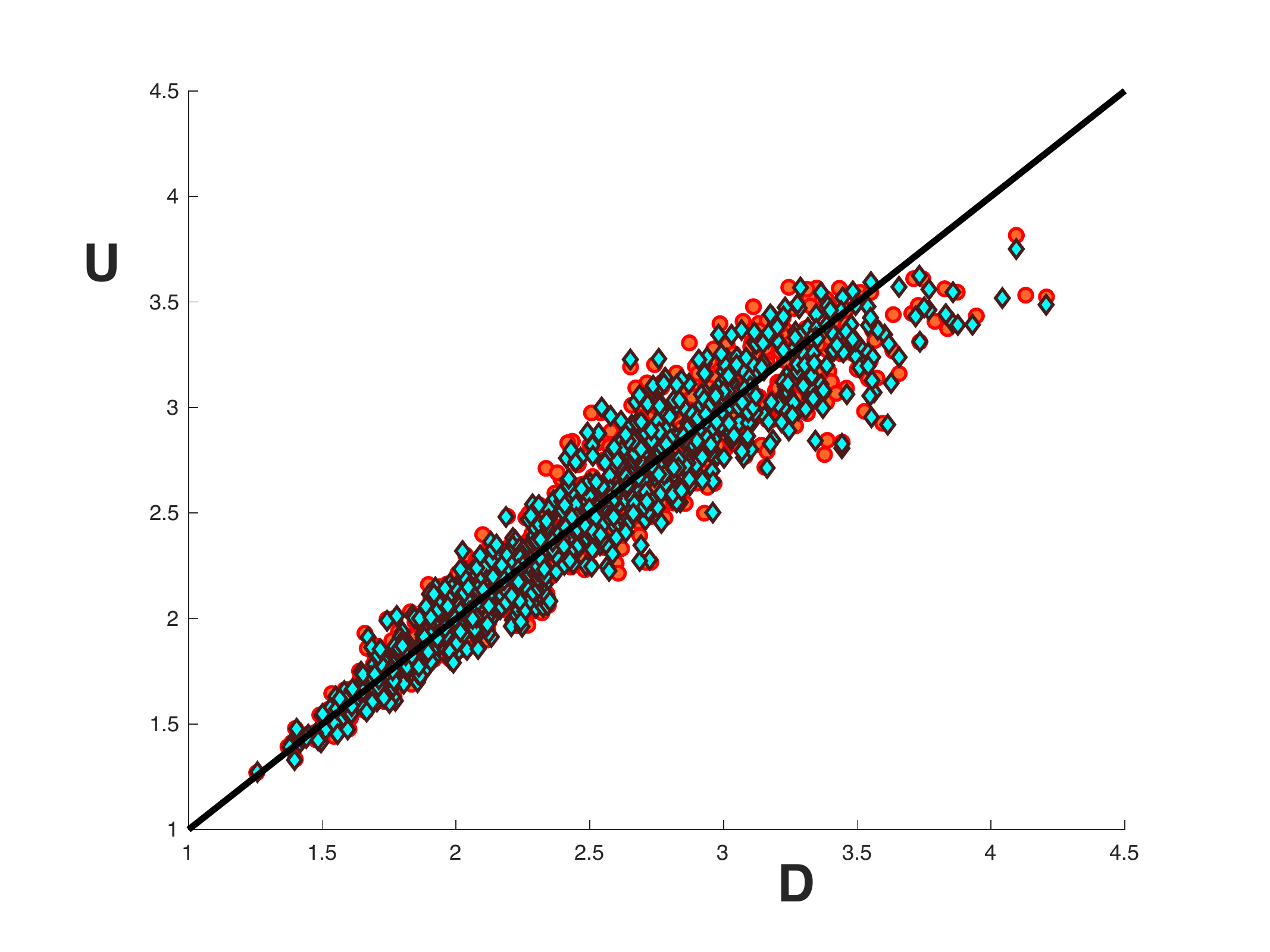}}
    \caption{Scatter plot between upstreamness and downstreamness of sector $i=7$ for the $N=200$ random model with exponential disorder with parameters $\mu=1$, $\mu_F=0.005$. Each of the $1000$ blue [red] points is obtained from one instance of the random ``source'' matrix $A$, and represents the pair of values ($(\bm U_1)_i,(\bm D_1)_i$) [($\tilde{U}_i,\tilde{D}_i$), respectively]. The thick black line has slope $+1$. }
    \label{fig:UvsD}
\end{figure}

\subsection{Model definition}\label{sec: modeldef}
In the random model we consider that the entries of $A$ are drawn independently from an exponential \textcolor{black}{pdf} $p(a)=\mu \exp(-\mu a)$ with mean $1/\mu$. As such, the entries of $A$ are all positive, and no economic or empirically motivated information whatsoever is injected in the construction of $A$. The final demand values $F_i$ $(i=1,\ldots,N)$ are further modeled as i.i.d. exponential random variables with mean $1/\mu_F$. \textcolor{black}{The choice of an exponential pdf for the entries of the random model instead of a fatter distribution, closer to the empirical one (see Section \ref{sec:reshuffling} below) makes the analytical calculations simpler and the interpretation of results more transparent. However, we tested numerically that the results remain robust even for other distributions (for which we do not provide an analytical treatment), therefore we can safely conclude that the choice of an exponential pdf is not in any way unduly restrictive or affecting the trustworthiness of our conclusions.}

From the matrix $A$, we construct the matrices $A_U$ and $A_D$ (see Eq. \eqref{eq:AU} and \eqref{eq:AD}, together with the definition of $Y_i$ in Eq. \eqref{eq:iteration1}) as
\begin{align}
(A_U)_{ij} &= \frac{a_{ij}}{\sum_j a_{ij}+F_i}\\
(A_D)_{ij} &= \frac{a_{ji}}{\sum_j a_{ij}+F_i}\label{ADdef}\ ,
\end{align}
where we used that $\sum_j a_{ij}+F_i = \sum_j a_{ji}+V_i$ for all $i$, as follows from the \textcolor{black}{input-output identities}.  Therefore, both $A_U$ and $A_D$ as defined above have non-negative elements, and are row sub-stochastic (as they should)\textcolor{black}{, provided that $\mu_F$ is sufficiently small. This condition $\mu_F\ll\mu$ is necessary to ensure that $F_i$'s will be (typically) large enough to make $\sum_j (A_D)_{ij}$ in \eqref{ADdef} smaller than $1$.}

For each instance of the random matrix $A$ and of the vector of final demands $\bm F$, we construct the matrices $A_U$ and $A_D$ as above, and from those we compute the pairs $(\bm U_1)_i,(\bm D_1)_i$ and $\tilde{U}_i,\tilde{D}_i$ for any sector $i$ that we choose. These are all random variables, whose pairwise covariance is of interest in this paper.

\section{Results}\label{sec:results}
Our results are summarized in the theorem and corollary below. We show that even in our completely random model (with no economic or empirically motivated information whatsoever injected in constructing the I/O table), the upstreamness and downstreamness of an industrial sector of a single country are necessarily positively correlated, and that for any $N$ the slope of the scatter plot between the two is always equal to $+1$.

In Figs. \ref{fig:UvsDlogn} and \ref{fig:UvsDunif} we further numerically check that our results do not crucially depend on the specific choice of the way the random matrices $A_U$ and $A_D$ are generated, so the positive correlation between upstreamness and downstreamness of economic sectors -- and their correlation slope being $+1$ -- seem to be very robust results and rather insensitive to the fine details of the inter-sectorial I/O matrix\textcolor{black}{, unless the matrix is exceedingly sparse and/or it represents a very \emph{ad hoc} structure (e.g. the linear chain of Section \ref{sec:curse}), which of course occur with zero likelihood in the random model}.

\begin{thm}
Let $N\times N$ matrices be defined as 
\begin{align}
(A_U)_{ij} &= \frac{a_{ij}}{\sum_j a_{ij}+F_i}\\
(A_D)_{ij} &= \frac{a_{ji}}{\sum_j a_{ij}+F_i}\ ,
\end{align}
where $a_{ij}$ are i.i.d. variables drawn from an exponential pdf $p(a)=\mu \exp(-\mu a)$, and the $F_i$'s are i.i.d. variables drawn from an exponential pdf $p_F(F)=\mu_F \exp(-\mu_F F)$ with $\mu_F\ll \mu$ to ensure that $A_U$ and $A_D$ are row sub-stochastic. Let $r=\sum_j (A_U)_{1j}$ and $r'=\sum_j (A_D)_{1j}$. Then the simplified covariance between upstreamness and downstreamness (see Eq. \eqref{eq:simplifiedcov})
\begin{equation}
    \mathcal{C}_N(\mu,\mu_F)=\frac{\mathbb{E}[rr']-\mathbb{E}[r]\mathbb{E}[r']}{(1-\mathbb{E}[r])(1-\mathbb{E}[r'])}
\end{equation}
is given by the exact formula $\mathcal{C}_N(\mu,\mu_F)=-\mathcal{N}_N(\phi)/\mathcal{D}_N(\phi)$, where

\begin{strip}
\begin{dgroup*}
\begin{dmath*}
\mathcal{N}_N(\phi) =   (\phi -1) \left[N (\phi -1) \mathrm{B}(1,N+1)^2 \, _2F_1(1,N+1;N+2;\phi ){}^2+(N-1) N (\phi -1) \mathrm{B}(1,N+1) \, _2F_1(1,N+1;N+2;\phi ) (\mathrm{B}(1,N) \, _2F_1(1,N;N+1;\phi )+1)+(N+1) (\phi -1) \mathrm{B}(1,N+2) \, _2F_1(1,N+2;N+3;\phi )+N\right]
\end{dmath*}
\begin{dmath*}
\mathcal{D}_N(\phi)=((N-1) (\phi -1) \mathrm{B}(1,N) \, _2F_1(1,N;N+1;\phi )+(\phi -1) \mathrm{B}(1,N+1) \, _2F_1(1,N+1;N+2;\phi )+1) (N (\phi -1) \mathrm{B}(1,N+1) \, _2F_1(1,N+1;N+2;\phi )+1)\ ,
\end{dmath*}
\end{dgroup*}
\end{strip}
where $\phi=1-\mu_F/\mu$. Here, $\mathrm{B}(x,y)=\Gamma(x)\Gamma(y)/\Gamma(x+y)$ is the Beta function, and $_2 F_1$ is the Gaussian hypergeometric function.
\label{thm1}
\end{thm}

\begin{cor}
In the hypotheses of Theorem \ref{thm1}, the slope $S(\mu,\mu_F)$ of the scatter plot between the rank-1 estimators of downstreamness and upstreamness (see Eq. \eqref{eq:approximateslope}) is equal to $+1$ for any $N$, irrespective of the values of $\mu,\mu_F$.\label{cor1}
\end{cor}

The proofs are deferred to the Appendix.

\section{Numerical simulations}
{\label{sec:NumericalSimulations}}

We have performed numerical simulations on our random model, generating $m$ instances of the $N\times N$ matrix $A$ with i.i.d. exponential entries with mean $1/\mu$. We also generate random vectors of final demands $\bm F$ of size $N$, with i.i.d. entries with mean $1/\mu_F$ (with $\mu_F\ll \mu$).

For each generated instance of the matrix $A$, we formed the matrices $A_U$ and $A_D$ (as defined in Eqs. \eqref{eq:AU} and \eqref{eq:AD}), which have by construction non-negative elements, and are row sub-stochastic\footnote{For $A_U$ substochasticity is guaranteed by construction. For $A_D$ this is true with overwhelming likelihood provided that $\mu_F\ll\mu$.}.

From the matrices $A_U$ and $A_D$ so generated, we constructed the vectors of upstreamness $\bm U_1$ and downstreamness $\bm D_1$ values according to the inversion formulae Eq. \eqref{eq:up1} and \eqref{eq:D1}, respectively. We then pick a certain sector index $i$ (for example, $i=7$), and for that index we compute the estimators $\tilde U_i$ and $\tilde D_i$ according to Eqs. \eqref{eq:tildeUU} and \eqref{eq:tildeDD} respectively.

We first show in Figs. \ref{fig:scatterU} and \ref{fig:scatterD} that the ``true'' upstreamness (downstreamness) of sector $i$ -- computed from the full inversion formulae -- is perfectly correlated (with correlation slope $=+1$) with its approximate estimator. It is therefore perfectly justified to use the approximate estimators (instead of the full definition) to study correlations, as those are much simpler to handle analytically.

Next, in Table \ref{table:cov}, we report values of the ``true'' covariance between upstreamness and downsteamness of sector $i=7$, obtained from averaging over $m=10000$ numerically generated instances of our random model, against the values of $\mathcal{C}_N(\mu,\mu_F)$ analytically computed, and we observe an excellent agreement between the two. 

In Fig. \ref{fig:UvsD} we further provide scatter plots of upstreamness vs. downstreamness of sector $i$ (both ``true'' and approximate) - where each generated instance contributes a single point to the scatter plot. Again, we observe an excellent collapse onto the diagonal line with slope $+1$, further confirming that a strong positive correlation between upstreamness and downstreamness of the same sector is a generic feature of ``structure-less'' matrices --  provided they have non-negative entries and are sub-stochastic.

\begin{figure}[h]
    \centering
    \fbox{\includegraphics[scale = 0.20]{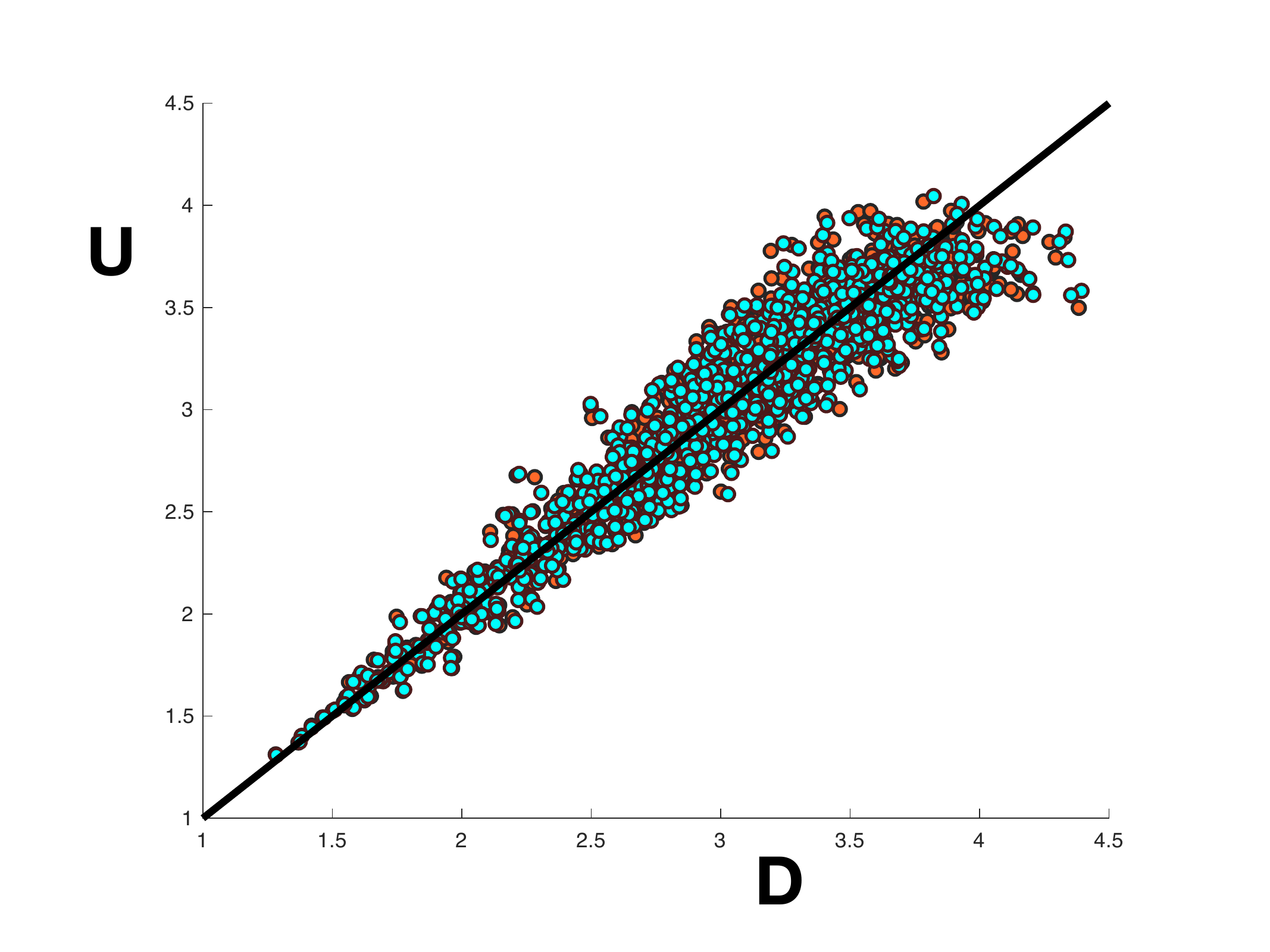}}
    \caption{Scatter plot between upstreamness and downstreamness of sector $i=7$ for the $N=400$ random model with log-normal disorder with parameters $\mu'=1$, $\mu_F'=6.67$. Each of the $1000$ light blue [orange] points is obtained from one instance of the random ``source'' matrix $A$, and represents the pair of values ($(\bm U_1)_i,(\bm D_1)_i$) [($\tilde{U}_i,\tilde{D}_i$), respectively]. The thick black line has slope $+1$. }
    \label{fig:UvsDlogn}
\end{figure}

\begin{figure}[h]
    \centering
    \fbox{\includegraphics[scale = 0.20]{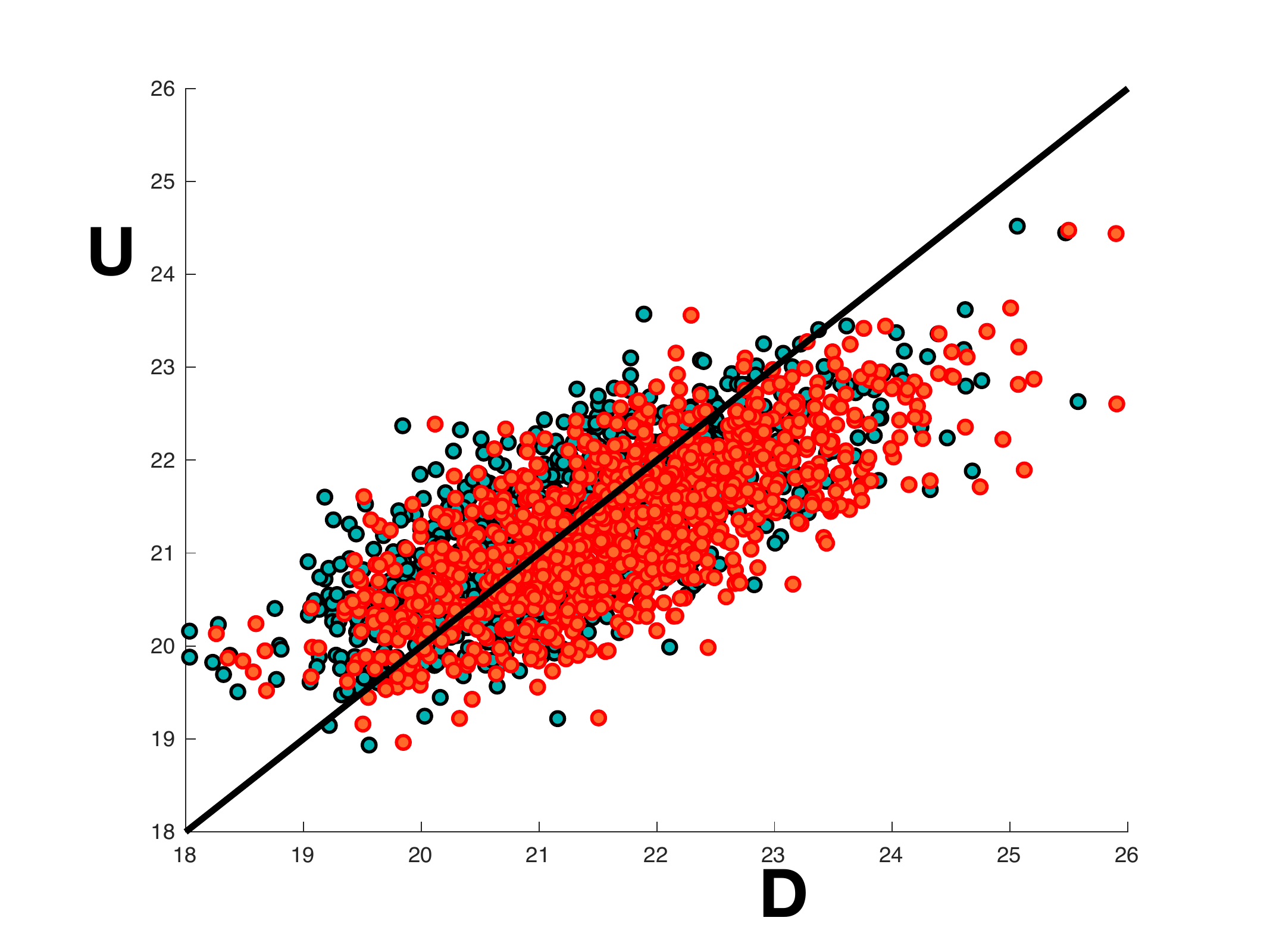}}
    \caption{Scatter plot between upstreamness and downstreamness of sector $i=7$ for the $N=400$ random model with uniform disorder with parameters $\mu=1$, $\mu_F=0.05$. Each of the $1000$ light blue [orange] points is obtained from one instance of the random ``source'' matrix $A$, and represents the pair of values ($(\bm U_1)_i,(\bm D_1)_i$) [($\tilde{U}_i,\tilde{D}_i$), respectively]. The thick black line has slope $+1$. }
    \label{fig:UvsDunif}
\end{figure}

\textcolor{black}{Moreover}, in Figs. \ref{fig:UvsDlogn} and \ref{fig:UvsDunif} we provide the same scatter plots as in Fig. \ref{fig:UvsD}, but this time for the ``original'' matrix $A$ (and similarly for the vector $\bm F$) having i.i.d. non-negative entries drawn from a log-normal ($p(a)=(a\sqrt{2\pi})^{-1} \exp (-(\ln(a)-\mu')^2/2)$) and uniform (with mean $1/\mu$ and $1/\mu_F$) pdf, respectively. Although we do not provide analytical results for these cases, these plots further confirm that the positive correlation with slope $+1$ between upstreamness and downstreamness keeps holding irrespective of the precise details of the way the ``source'' matrix $A$ is generated -- provided that $A_U$ and $A_D$ are non-negative and sub-stochastic. 

\begin{table}[ht]
\caption{Covariance between Upstreamness and Downstreamness in the random model for $i=7$ and taken over $m=10000$ samples} 
\centering 
\begin{tabular}{c c c c c} 
\hline\hline 
$\mu$ & $\mu_F$ & $N$ & $\mathrm{Cov}((\bm U_1)_i,(\bm D_1)_i)$ & $\mathcal{C}_N(\mu,\mu_F)$ \\ [0.5ex] 
\hline 
1 & 0.001 & 200 & 0.10450 & 0.10385 \\ 
2 & 0.005 & 400 & 0.30415 & 0.29494 \\
3 & 0.001 & 300 & 0.06136 & 0.06158\\
1.2 & 0.001 & 500 &  0.17346  & 0.17260\\
1.5 & 0.003 & 350 &  0.24224 & 0.23955\\ [1ex] 
\hline 
\end{tabular}
\label{table:cov} 
\end{table}

\textcolor{black}{In other experiments, we have further tested the effect of (random) ``sparsification'' of the matrix $A$, where elements of the random matrix $A$ were set to zero with a certain probability $1-p$. Computing the Pearson correlation coefficient between U and D of the same sector as a function of the sparsity parameter $p$ for the ``sparsified'' exponential model described above, we find that the correlation between U and D survives at $\sim 90\%$ level -- unless the sparsity increases to unrealistic values $(p\sim 0.2)$ -- and remains well above $\sim 60\%$ for a sparsity level as high as $p\sim 0.05$, where the matrix $A$ is essentially entirely dominated by zeros and it is not even guaranteed that the corresponding GVC remains connected (see Fig. \ref{fig:pearson} below). This finding is consistent with the ``chain'' geometry of Section \ref{sec:curse}: decreasing the sparsity of the chain, up to the extreme case of a fully connected economy, has generally the effect of increasing the U-D correlation.}

\begin{figure}[h]
    \centering
    \includegraphics[width=0.5\textwidth]{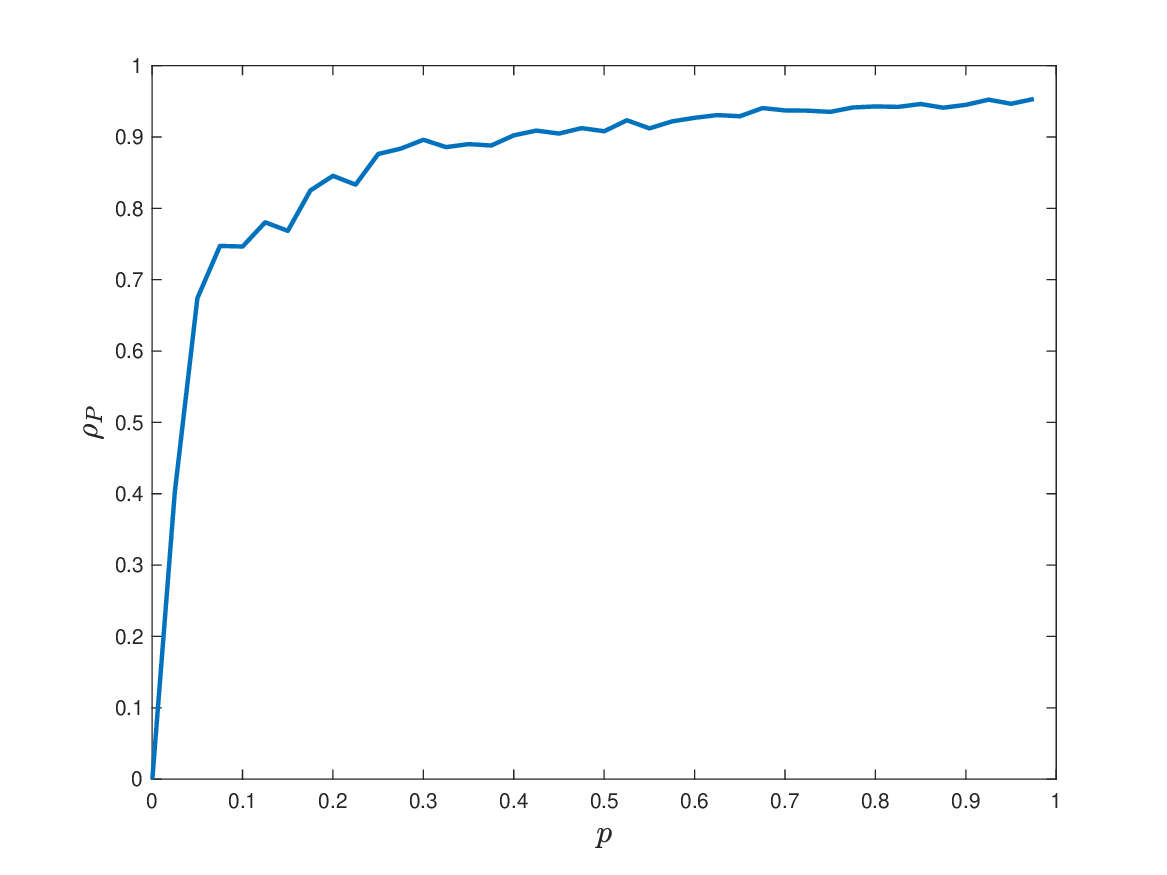}
    \caption{\textcolor{black}{For the ``sparsified'' exponential model, Pearson correlation coefficient between U and D of the same sector as a function of the sparsity parameter $p$ (matrix $A$ gets sparser as $p$ gets lower).}}
    \label{fig:pearson}
\end{figure}
\textcolor{black}{Our analyses so far have focused on the properties of an individual economy. To test whether our results hold more generally, we extended our model to consider the simple case of two interacting countries, represented by a $2 \times 2$ block matrix. The diagonal blocks represent interactions within sectors of the same economy, while the off-diagonal blocks represent interactions between sectors across the two countries.}

\textcolor{black}{We assume the diagonal blocks are fully populated, while in the off-diagonal blocks, each link is present with probability $p$, similar to the sparse case considered earlier.}

\textcolor{black}{The two extreme cases of $p = 0$ and $p = 1$ correspond to two isolated economies and a `fully globalized' economy, respectively. The results we discussed above apply to these extreme cases: when $p = 1$, the system behaves as a single, larger economy, whereas for $p = 0$, each country can be analyzed individually.
}

\begin{figure}[h]
    \centering
    \includegraphics[width=0.5\textwidth]{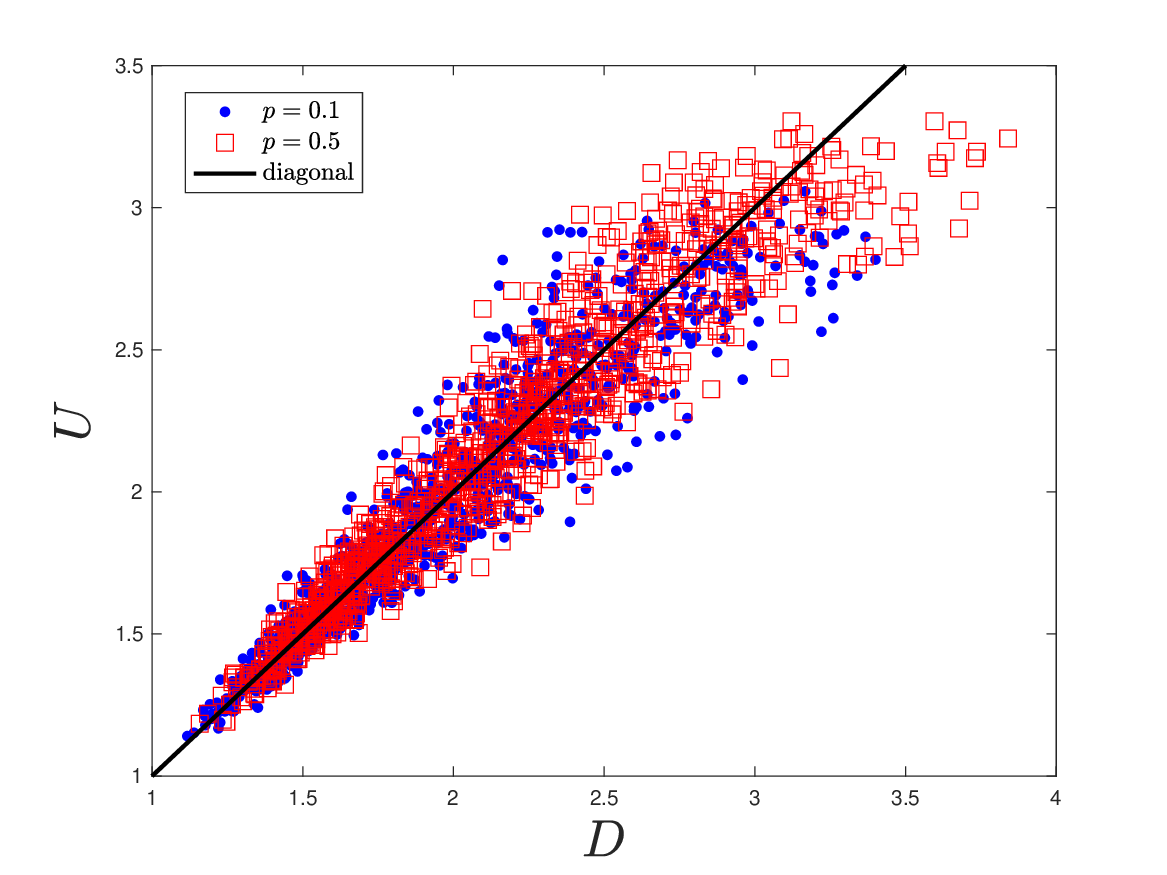}
    \caption{\textcolor{black}{Scatter plot between upstreamness and downstreamness of sector $i=7$ for the $2\times 2$ block model. The aggregate matrix is of size $N=200$, with each block of size $100\times 100$. Blue dots refer to $p=0.1$, while red squares to $p=0.5$ for the off-diagonal blocks. The diagonal blocks are fully populated, and the non-zero entries of the matrix are drawn from an exponential distribution with parameter $\mu=1$, while $\mu_F=0.05$.   The solid black line has slope $+1$. }}
    \label{fig:blockEconomy}
\end{figure}

\textcolor{black}{By tuning $p$, we can observe the effect of adjusting the level of integration between the economies, for instance, through trade barriers, as discussed in \cite{Antras2018}. As shown in Figure \ref{fig:blockEconomy}, where we present scatter plots of downstreamness and upstreamness for different values of $p$, we find that the model's qualitative behavior and the ensuing U-D correlations remain largely independent of $p$.}

\section{Random reshuffling of the I-O Table}\label{sec:reshuffling}

\begin{figure}
    \centering
    \begin{subfigure}[t]{0.49\textwidth}
    \centering
    \includegraphics[width=\textwidth]{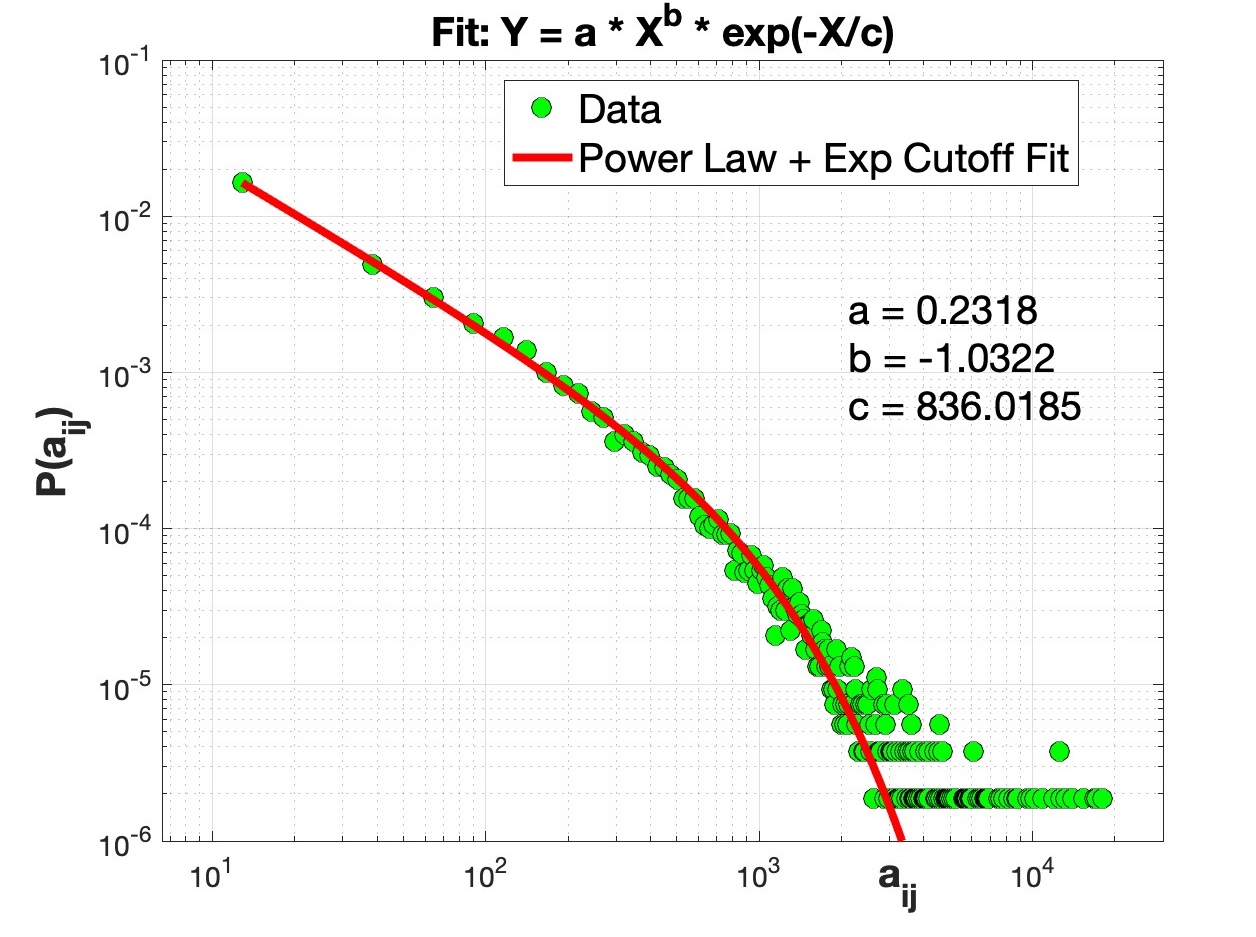}
    \caption{Poland} \label{fig:Poland}
     \end{subfigure}
    \hfill
    \begin{subfigure}[t]{0.49\textwidth}
    \centering
    \includegraphics[width=\textwidth]{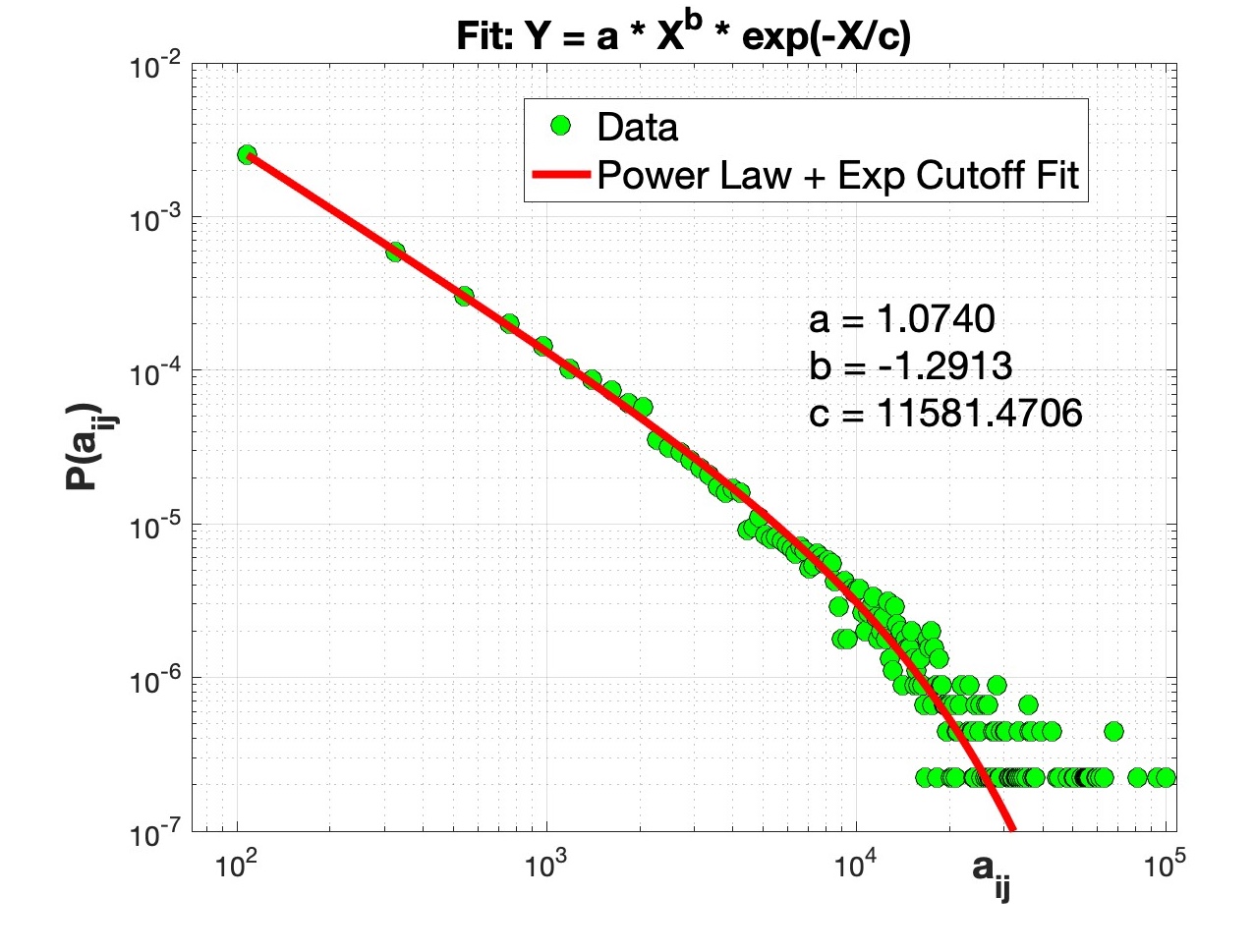}
    \caption{France} \label{fig:France}
     \end{subfigure}
    \hfill
    \begin{subfigure}[t]{0.49\textwidth}
    \centering
    \includegraphics[width=\textwidth]{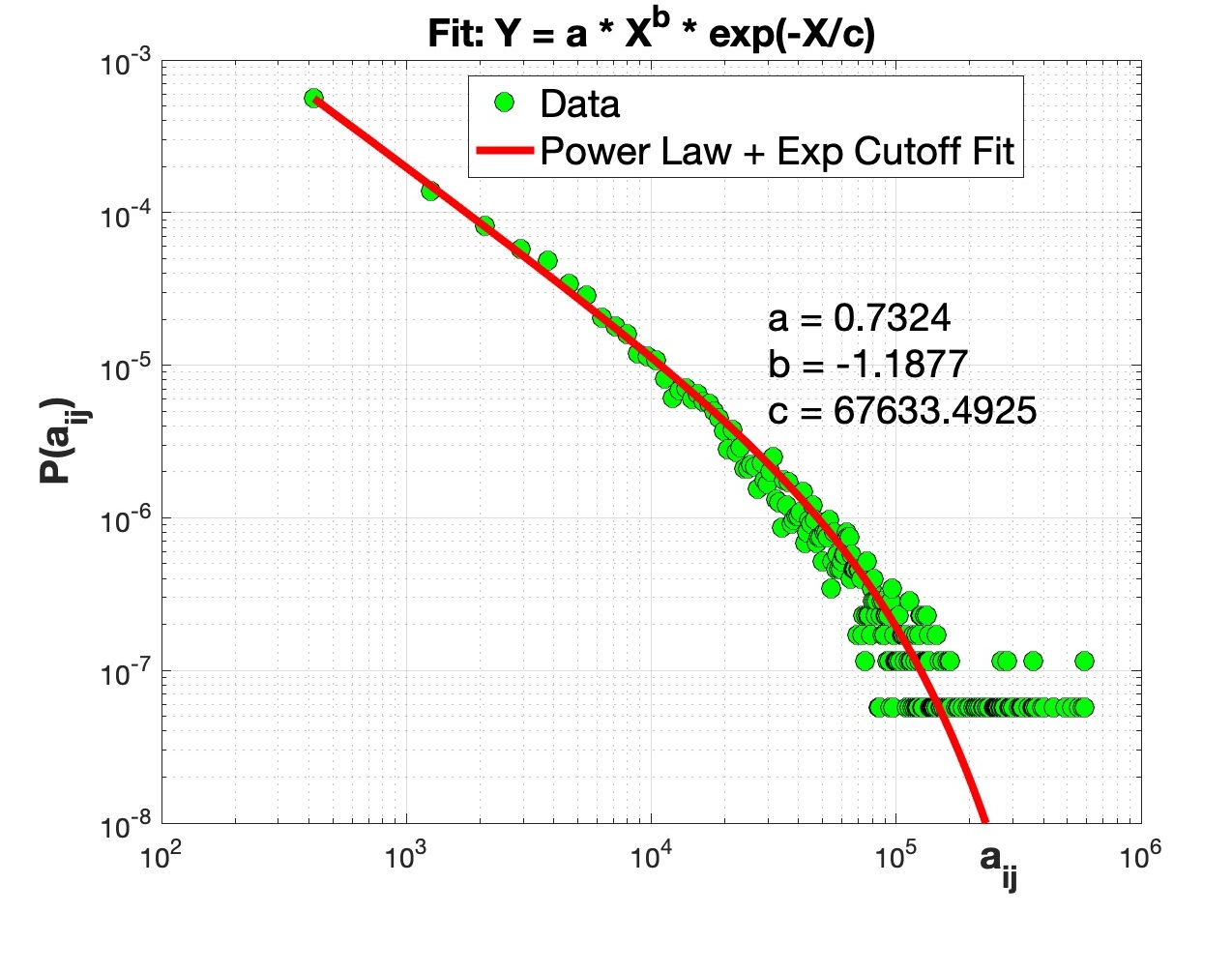}
    \caption{U.S.A.} \label{fig:USA}
     \end{subfigure}
    \hfill
     \caption{\textcolor{black}{Probability density function in log-log scale of the entries of WIOD matrices for three countries, obtained by histogramming all the entries of the I-O matrices across the available 16-years span. In red, the line of best fit with a power-law with exponential cutoff.}}
    \label{fig:WIODempirical}
\end{figure}

To perform the second series of experiments, we have taken the empirical I-O matrices including 39 countries for the years 1995-2011 (from WIOD 2013 release). 
\textcolor{black}{First, in Fig. \ref{fig:WIODempirical} we plot the empirical pdf of the entries of I-O matrices for three representative countries (Poland, France, and U.S.A.) across all available years. The empirical distributions are generally well-fitted by power laws with exponential cutoffs ($Y=a X^b \exp(-X/c)$), with best fitting parameters included inside the plot boxes. The essential power-law (fatter than exponential) nature of the entry distribution is partly due to a genuine ``multi-scale'' nature of the typical interactions among sectors, which span several orders of magnitude, but it is also partly affected by having to lump together matrices corresponding to different years to obtain a sufficiently meaningful statistics. 
}
\textcolor{black}{Next, }for each country and each year, we have computed the upstreamness and downstreamness of that country using Eq. \eqref{eq:fally2} and Eq. \eqref{eq:D1} respectively, averaging over sectors. 

\begin{figure}[h]
    \centering
    \fbox{\includegraphics[scale = 0.1]{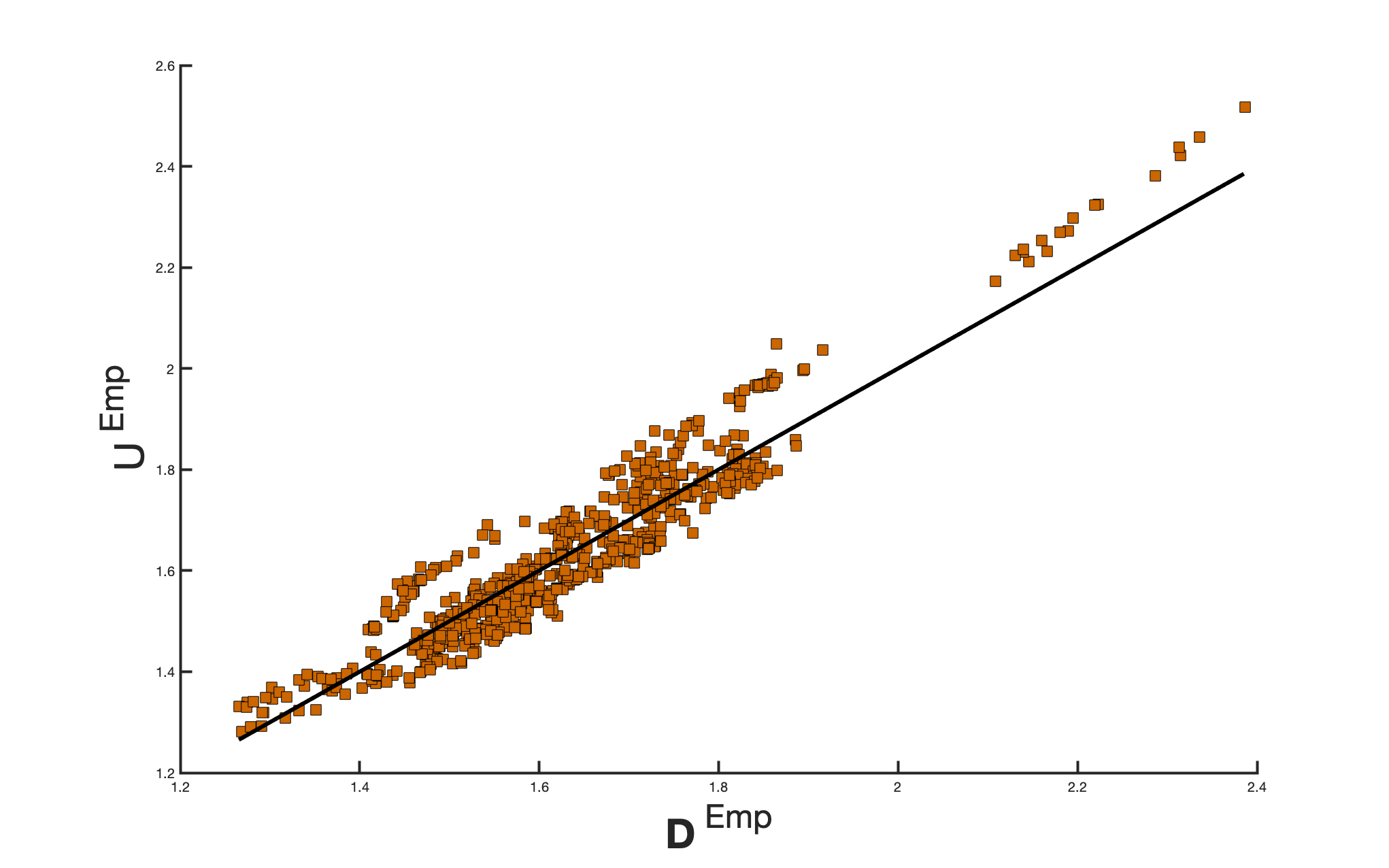}}
    \caption{Empirical upstreamness vs. empirical downstreamness (averaged over $35$ sectors) for 39 countries for the years 1995-2001.}
    \label{fig:UpDownAll}
\end{figure}

\begin{figure}[h]
    \centering
    \fbox{\includegraphics[scale = 0.12]{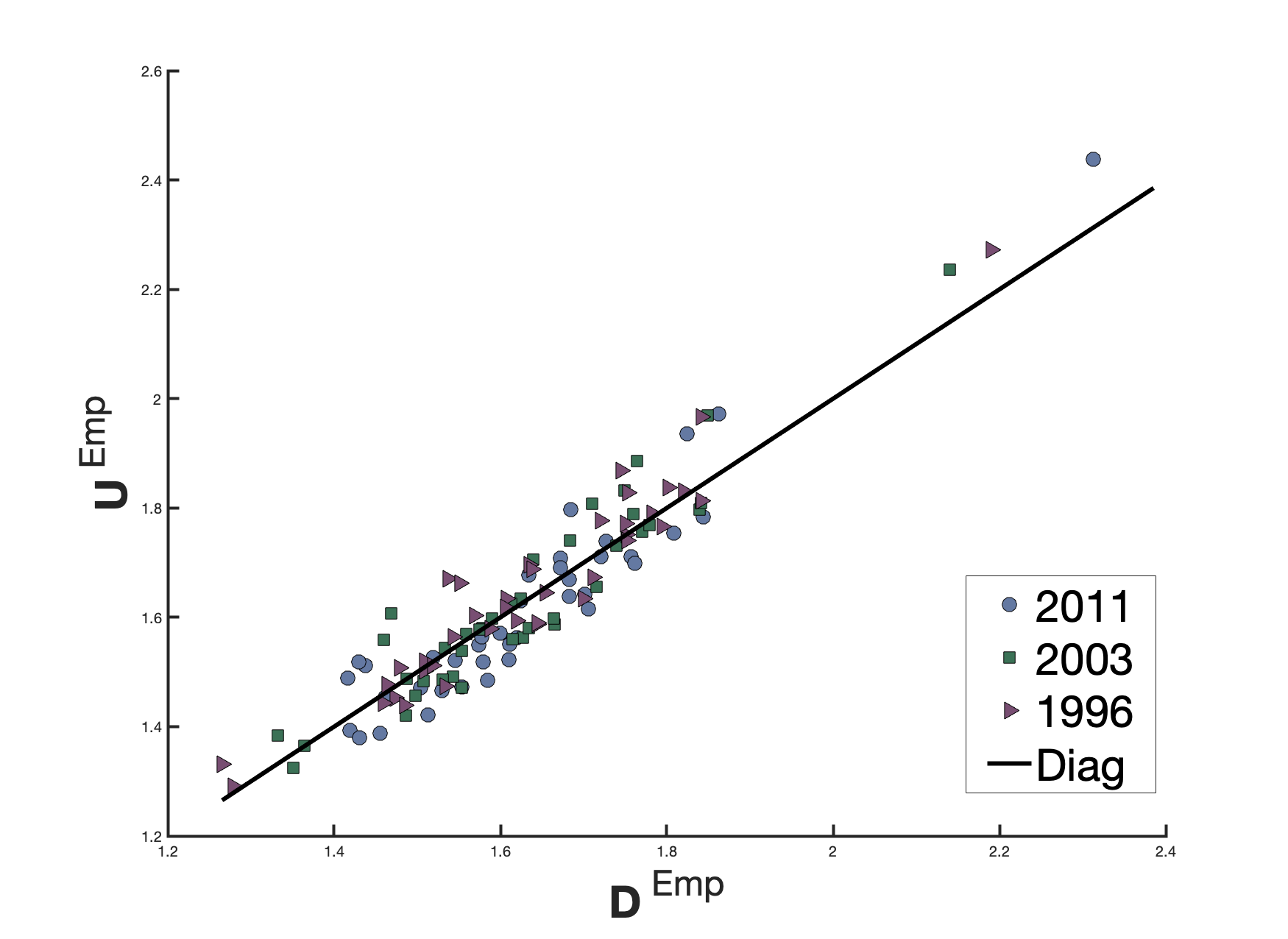}}
    \caption{Empirical upstreamness vs. empirical downstreamness (averaged over $35$ sectors) for 39 countries in selected years (1996-2003-2011).}
    \label{fig:UpDownSelected}
\end{figure}

In fig. \ref{fig:UpDownAll}, we show the values of upstreamness for all countries in all years, spanning a period identical to that considered in the paper \cite{Antras2018}. In Fig. \ref{fig:UpDownSelected}, we provide the scatter plot of upstreamness vs. downstreamness of each country for three selected years (1996-2003-2011).  As expected, we confirm the general trend observed in \cite{Antras2018}, namely that the two measures appear to be strongly correlated with a slope of the scatter plot very close to $+1$. 

\begin{figure}[h]
    \centering
    \fbox{\includegraphics[scale = 0.12]{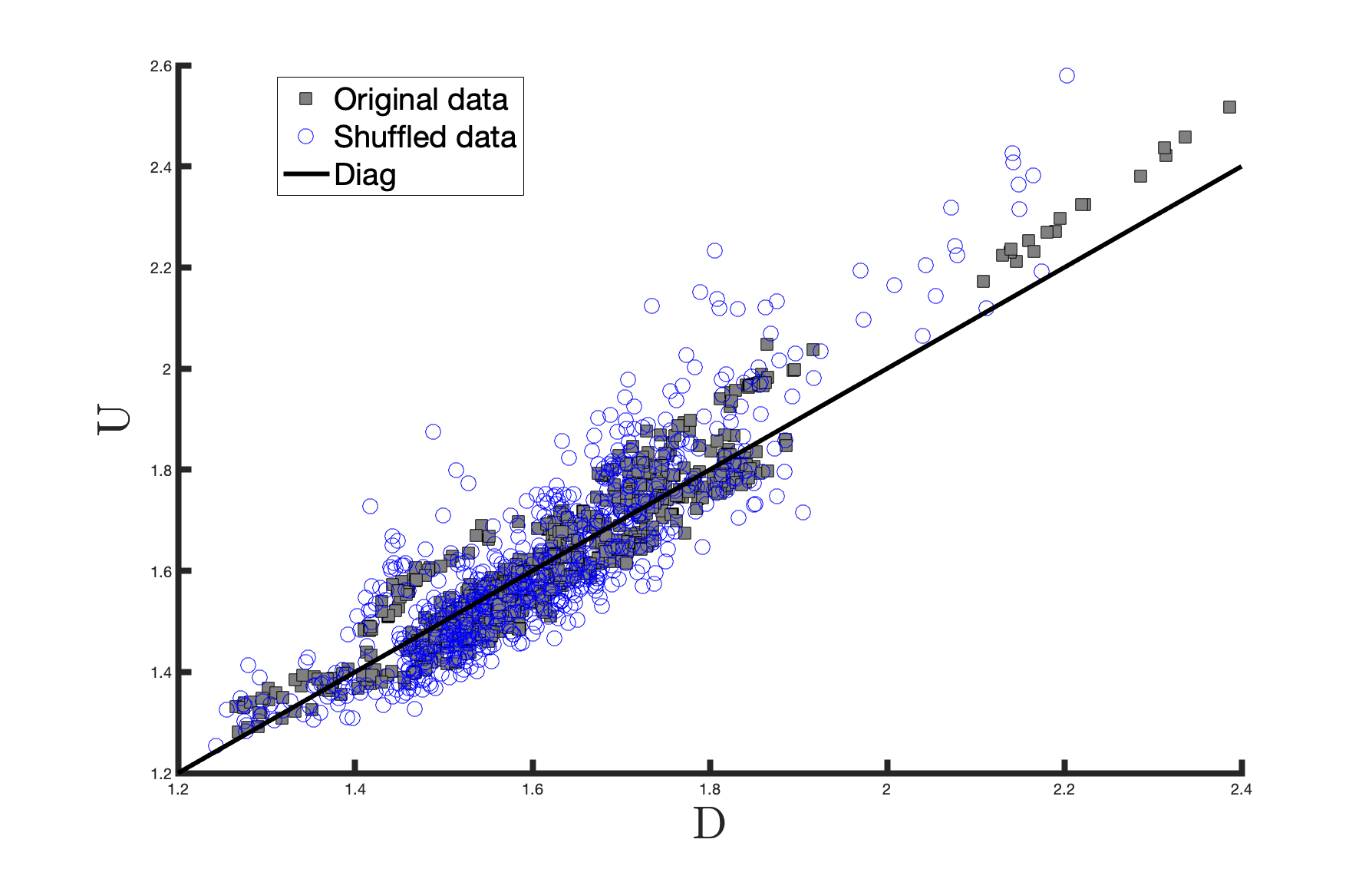}}
    \caption{Upstreamness vs. downstreamness (averaged over $35$ sectors) calculated on the empirical and reshuffled matrices for 39 countries for the years 1995-2001. Blue circles represent values calculated on I-O matrices where columns have been randomly reshuffled. Grey squares are upstreamness/donwstreamness pairs calculated on the original data.}
    \label{fig:shuffleallvsreal}
\end{figure}

In Fig. \ref{fig:shuffleallvsreal}, though, we took the same I-O matrices for the entire period 1995-2011 and we randomly reshuffled the columns of such matrices according to a random permutation of the set $\{1,\ldots,N\}$. The resulting matrices satisfy the same aggregate constraints (namely, the row sums) of the orginal, actual matrices, but the interactions between sectors have been randomly scrambled, resulting in an entirely fictitious Global Value Chain, where all economic forces at play in the real world have been neutralized. Still, and quite surprisingly, we find that the same linear correlation with slope close to $+1$ between upstreamness and downstreamness survives. We have checked that this result is not an artifact of the specific random permutation of columns chosen, but keeps holding irrespective of what new ``strength'' of interaction is attributed to pairs of sectors/countries via a random reshuffling of the old, actual one. 

The fact that U-D correlations are so strong and stable that they survive a complete overhauling of the actual economic interactions at play in the real world provides a further strong confirmation that most -- if not all -- of such correlations cannot be due to sophisticated and finely-tuned economic factors leading to a \emph{specific} set of inter-sectorial interactions, otherwise any random reshuffling would have completely annihilated them. These experiments therefore lend further support to the claim that U-D correlations are mostly due to structural constraints that the matrices $A_U$ and $A_D$ must meet simply because of the way they are constructed from the interaction matrix $A$.

\textcolor{black}{While our work provides an intuition on the origin of correlations between the measures of upsptreamness and downstreamness introduced in the literature, the challenge of how to} devise a better measure of downstreamness that \textcolor{black}{(i)} is \textcolor{black}{not trivially correlated to} the upstreamness\textcolor{black}{, and (ii) complies with the more intuitive notion of measuring the positioning of a sector along the production chain has been thoroughly addressed in \cite{criticaAntras}}, to which we refer for details (see also \cite{branger2023stock}).

\section{Discussion}\label{sec:discussion}

In summary, we have considered two classes of \textit{random} Input/Output matrices $A$ to test whether the ``puzzling'' correlation detected between upstreamness and downstreamness at the sector and country level \cite{Antras2018} would survive even if the underlying economic forces and the inter-sectorial dependencies had nothing to do with the real world.

First, we constructed a random model for the matrix $A$ that mimics a closed economy composed of $N$ economic sectors. We showed analytically that the resulting upstreamness and downstreamness of a given sector are generically positively correlated, with a slope of the scatter plot between the two equal to $+1$, if the entries of the matrix $A$ are independently drawn from an exponential pdf. We also showed numerically that our results do not depend very strongly on the pdf of matrix entries. At least at the level of single countries, our work provides a comforting ``proof of principle'' that a strong positive correlation between upstreamness and downstreamness of individual sectors as originally defined (see Eqs. \eqref{eq:up1} and \eqref{eq:D1}) is bound to materialize even on a structure-less and zero-information I-O matrix: one would have to try very hard to concoct an I-O matrix so extraordinary and finely tuned, that such correlations were \emph{not} observed. 

Secondly, we started from a real, empirical I-O matrix $A$ taken from the WIOD Dataset (2013 Release), which displayed the same kind of correlations between observables as originally detected in \cite{Antras2018}. We performed several experiments where we simply randomly reshuffled the columns of the interaction matrix prior to computing the matrices $A_U$ and $A_D$ from which upstreamness and downsteamness of the same sector can be determined. The resulting shuffled interaction matrix $A'$ is a new, perfectly legitimate interaction matrix, which shares the row sums and all other structural constraints with the original matrix $A$, but whose economic fabric and inter-sectorial dependencies are entirely made up: the Global Value Chain that $A'$ embodies does not respond to any realistic economic force nor is connected to any realistic economic scenario. Yet, we find that the correlations survive unscathed. 

\textcolor{black}{By further analyzing a simplified geometry for the economy (i.e. a linear chain) in Section \ref{sec:curse}, we identified further subtleties with the standard definitions of upstreamness and downstreamness, which are jointly unable to support -- even in such a simple geometry -- the intuitive interpretation in terms of positioning of a sector along the chain with respect to final demand and primary factors of production, respectively. We were able to trace such less intuitive features of the U-D measures to the \textcolor{black}{input-output identities} that any I-O matrix must satisfy. This ``curse of the \textcolor{black}{input-output identities}'' effectively introduces additional links into the I-O table as extra value must be added at each node, which in turn leads to uncontrolled effects on the numerical values taken by the U and D indicator, blurring their natural economic interpretations and pushing one indicator to essentially align with the other (with the exclusion of special cases constructed \emph{ad hoc}, e.g. unphysical levels of sparsity, or vanishing value added at each node)}.

\textcolor{black}{Although derived in the context of a closed-economy I-O table (see Fig. \ref{fig:ioscheme2}), our results are nevertheless also relevant to the more general setting of international trade considered in \cite{Antras2018} and the puzzling correlation highlighted thereof for the following reasons: (i) the ``giant'' I-O matrix that includes inter-country trade blocks satisfies the same constraints (non-negative entries, and row-stochasticity) as the single-country one (and thus as well as the random model we presented). (ii) The model considered here -- after a trivial re-interpretation of the matrix entries -- is at the very least expected to mimic rather accurately what would happen in the more general inter-country setting in the two extreme cases of \emph{zero} and \emph{infinite} trade barriers between different countries. In the former case, the ``giant'' I-O matrix will have inter-country and intra-country blocks that do not differ much (statistically), therefore -- after country-wise aggregation -- the resulting $A$ matrix will look very similar to the one we considered in this paper. In the latter case, the ``giant'' I-O matrix will be block-diagonal -- with inter-country blocks full of zeros due to the absence of trade -- with each non-zero (intra-country) block being an independent replica of the closed economy model proposed here.} 

\textcolor{black}{In fact, we carried out simulations for the case when the random matrix $A$  has a $2\times 2$ block structure to mimic the case of two interacting countries. We considered fully dense diagonal blocks to model interactions of the sectors within each country, and sparse off-diagonal blocks to model interactions between countries. In this simple model, the sparsity of off-diagonal blocks can be changed to study the effect of different levels of integration between the countries. We found that the positive correlation between upstreamness and downstreamness we documented for the close economy persists for this more structured economy, pretty much irrespective of the level of integration among the two countries of our simplified setting.}

While a deeper investigation of the intermediate trade barriers setting in the random ``giant'' model is surely needed, the aforementioned observations sharply point towards the observed correlation between upstreamness and downstreamness also at country level being simply due to structural and unavoidable algebraic constraints that I-O tables and their surrogates must satisfy.

Our first series of results rest on the following assumptions and simplifications:
\begin{enumerate}
    \item The correlation between ``true'' upstreamness and downstreamness of a sector can be faithfully probed by using the rank-$1$ approximants defined in \cite{Bartolucci,Bartolucci2}. This assumption was tested on empirical I/O data in \cite{Bartolucci,Bartolucci2}, and on the random model here in Figs. \ref{fig:scatterU} and \ref{fig:scatterD}, by showing that the ``true'' upstreamness (or downstreamness) is indeed perfectly correlated with its rank-$1$ approximant. Such rank-$1$ approximation could only become less reliable if the true Input/Output matrix were exceedingly sparse, i.e. with a very large number of zero entries (see discussion in \cite{Bartolucci2,gap1,gap2}), a situation that does not often materialize in practice. By considering national I-O tables available from the 2013 release of the WIOD \cite{wiotdataset}, we indeed obtain quite high average densities of nonzero elements -- between 0.92 and 0.93 across 40 countries for the years 1995-2011. We have further checked that a moderate sparsification of our random model does not qualitatively change our conclusions, however in future experiments it will be appropriate to test the consequences of sparsity more thoroughly.  
    \item We have assumed that the entries of the matrix $A$ were independent and identically distributed (i.i.d.). Some preliminary results (not shown) where this assumption has been relaxed indicate that heterogeneity in the pdfs of the entries of $A$ may not play a major role and is generally insufficient to change the conclusions of our analysis.  
    \item We used some simplifications (for instance, appealing to the LLN) to make some progress in the analytical calculations. All approximations are controlled and have been carefully tested.
\end{enumerate}

Apart from performing a more thorough study on the effect of sparsity and heterogeneity in random models of I-O tables, in future studies it will be interesting to try to compute analytically the \emph{full} covariance Eq. \eqref{eq:full_covariance} for our random model and for various different pdfs of the entries of the I/O matrix $A$, i.e. without employing any rank-$1$ proxy and/or LLN approximations. This task will require handling the average of (products of) inverse matrices (coming from the definitions of upstreamness and downstreamness, see Eq. \eqref{eq:up1} and \eqref{eq:D1}), which is possible in some cases using techniques from statistical physics \cite{Bartolucci}.

\appendix

\section{Derivation of Theorem \ref{thm1} and Corollary \ref{cor1}}
\label{sec:sample:appendix1}

We need to compute $\mathbb{E}[r]$, $\mathbb{E}[r']$ and $\mathbb{E}[rr']$ separately. We have
\begin{equation}
    \mathbb{E}[r]=\int_0^\infty \prod_{i=1}^N da_i p(a_i)\int_0^\infty dF~p_F(F)\frac{\sum_k a_k}{\sum_k a_k + F}\ ,
\end{equation}
where for simplicity we denoted $a_k \equiv a_{1k}$ and $F\equiv F_1$. Using the identity
\begin{equation}
    \frac{1}{\xi}=\int_0^\infty ds~e^{-\xi s}\qquad \xi>0\label{eq:identity}
\end{equation}
we have
\begin{align}
\nonumber\mathbb{E}[r] &=\mu^N\mu_F\int_0^\infty \prod_{i=1}^N da_i\int_0^\infty dFds~\sum_k a_k  \times\\
\nonumber &\times e^{-\mu\sum_k a_k-\mu_F F-s(\sum_k a_k+F)}\\
\nonumber &=\mu^N\mu_F N \int_0^\infty ds \left[\int_0^\infty da e^{-\mu a-s a}\right]^{N-1}\times\\
\nonumber &\times\int_0^\infty dy~y e^{-\mu y-s y}\int_0^\infty dF e^{-\mu_F F-s F}\\
&=\boxed{\mu^N\mu_F N J(N+1)}\label{eq:avR}
\end{align}

where
\begin{align}
\nonumber J(k) &= \int_0^\infty ds \frac{1}{(\mu+s)^{k}}\frac{1}{\mu_F+s}=\\ \nonumber &=\frac{1}{\mu^k}\int_0^\infty dt (1+t)^{-1} \left(1+\frac{\mu_F}{\mu}t\right)^{-k}=\\
&=\frac{1}{\mu^k}\mathrm{B}(1,k)~ _2F_1(k,1;k+1;1-\mu_F/\mu)
\end{align}
from \cite{Grad}, formula 3.197.5 (pag. 335) with $\lambda=1$, $\nu=-1$, $\alpha=\mu_F/\mu$ and $\tilde\mu=-k$. Here, $\mathrm{B}(\cdot,\cdot)$ is the Beta function, and $_2 F_1$ is a hypergeometric function.

Similarly
\begin{align}
 \nonumber   \mathbb{E} &[r']=\mu^{2N-1}\mu_F\int_0^\infty \prod_{i=1}^N da_i p(a_i) \prod_{j=2}^N db_j p(b_j)\times\\
\nonumber    &\times\int_0^\infty dF~p_F(F)\frac{a_1+\sum_{k\geq 2}b_k}{\sum_k a_k+F}\\
    &=\boxed{\mu^{2N-1}\mu_F \left[\frac{1}{\mu^{N-1}}J(N+1)+\frac{N-1}{\mu^N}J(N)\right]}\ ,\label{expectrprime}
\end{align}
where for simplicity we denoted $a_{1j}\equiv a_j$ (for $j=1,\ldots,N$), and $a_{k1}\equiv b_k$ (for $k=2,\ldots,N$). To prove this, we write
\begin{equation}
 \mathbb{E}[r']=\mu^{2N-1}\mu_F [I_1+I_2]\ ,   
\end{equation}
where

\begin{align}
\nonumber I_1 &=\int_0^\infty ds \left[\int_0^\infty dx e^{-\mu x-s x}\right]^{N-1}\int_0^\infty dy~y e^{-\mu y-s y}\\
\nonumber &\times \left[\int_0^\infty dz~e^{-\mu z}\right]^{N-1}\int_0^\infty dF e^{-\mu_F F-s F}\\
&=\frac{1}{\mu^{N-1}}\int_0^\infty ds\frac{1}{(\mu+s)^{N+1}(\mu_F+s)}=\boxed{\frac{J(N+1)}{\mu^{N-1}}}
\end{align}

and

\begin{align}
\nonumber I_2 &=(N-1)\int_0^\infty ds \left[\int_0^\infty dx~ e^{-\mu x-sx}\right]^N \times\\ \nonumber &\times \int_0^\infty dF e^{-\mu_F F-s F}\left[\int_0^\infty dy e^{-\mu y}\right]^{N-2}\times \\
\nonumber &\int_0^\infty dz~z e^{-\mu z}=\frac{N-1}{\mu^N}\int_0^\infty ds \frac{1}{(\mu+s)^N (\mu_F+s)}\\
\nonumber &=\boxed{\frac{N-1}{\mu^N}J(N)}\ .
\end{align}

Finally
\begin{align}
   \nonumber &\mathbb{E}[rr'] =\mu^{2N-1} \mu_F \int_0^\infty \prod_{i=1}^N da_i p(a_i)\prod_{j=2}^N db_j p(b_j)dF\\
 \nonumber   &\frac{\sum_\ell a_\ell}{\sum_k a_k+F}\frac{a_1+\sum_{k\geq 2}b_k}{\sum_k a_k+F}=\\
 \nonumber & =\mu^{2N-1}\mu_F \left[\frac{2}{\mu^{N-1}}L(N+2)+\frac{N-1}{\mu^{N}}L(N+1)\right.\\
  &\left. +\frac{N-1}{\mu^{N-1}}L(N+2)+\frac{(N-1)^2}{\mu^{N}}L(N+1) \right]\ ,\label{eq:rrprime}
\end{align}
where 
\begin{align}
\nonumber L(k)&=\int_0^\infty ds dt \frac{1}{\mu_F+s+t}\frac{1}{(\mu+s+t)^{k}}=\\
&=\frac{\mu^{1-k}}{k-1}-\mu_F J(k)\ .
\end{align}

Eq. \eqref{eq:rrprime} follows from writing $\sum_\ell a_\ell = a_1 +\sum_{\ell\neq 1} a_\ell$, and applying the ``lifting-up'' identity \eqref{eq:identity} twice, which yields
\begin{equation}
   \mathbb{E}[rr'] =\mu^{2N-1} \mu_F[K_1+K_2+K_3+K_4]\ ,
\end{equation}
where
\begin{align}
\nonumber K_1 &= \int dx_1\cdots dx_N dF_1 dy_2\cdots dy_N ds dt\\
\nonumber & e^{-\mu \sum_k x_k -\mu\sum_{k\geq 2}y_k-\mu_F F_1}x_i^2\times\\
\nonumber &\times e^{-s(\sum_k x_k+F_1)-t(\sum_k x_k+F_1)}=\\
\nonumber &=\frac{1}{\mu^{N-1}}\int ds dt \frac{1}{\mu_F+s+t}\left[\int dx~ e^{-\mu x-(s+t)x}\right]^{N-1}\\
\nonumber &\times\int dx~x^2 e^{-\mu x - (s+t)x}\\
\nonumber &=\frac{2}{\mu^{N-1}}\int_0^\infty ds dt \frac{1}{\mu_F+s+t}\frac{1}{(\mu+s+t)^{N+2}}=\\
&=\boxed{\frac{2}{\mu^{N-1}}L(N+2)}\ .
\end{align}

\begin{align}
\nonumber K_2 &= \int dx_1\cdots dx_N dF_1 dy_2\cdots dy_N ds dt\\
\nonumber &\times e^{-\mu \sum_k x_k -\mu\sum_{k\geq 2}y_k-\mu_F F_1}x_i\sum_{k\geq 2}y_k\times\\
\nonumber &\times e^{-s(\sum_k x_k+F_1)-t(\sum_k x_k+F_1)}=\\
\nonumber &=(N-1)\int ds dt \frac{1}{\mu_F+s+t}\left[\int dx~ e^{-\mu x-(s+t)x}\right]^{N-1}\\
\nonumber &\times\int dx~x e^{-\mu x - (s+t)x}\\
\nonumber &\times \left[\int dy e^{-\mu y}\right]^{N-2}\int dy~y e^{-\mu y}\\
\nonumber &=\frac{N-1}{\mu^{N}}\int_0^\infty ds dt \frac{1}{\mu_F+s+t}\frac{1}{(\mu+s+t)^{N+1}}=\\
&=\boxed{\frac{N-1}{\mu^{N}}L(N+1)}\ .
\end{align}

\begin{align}
\nonumber K_3 &= \int dx_1\cdots dx_N dF_1 dy_2\cdots dy_N ds dt\\
\nonumber &\times e^{-\mu \sum_k x_k -\mu\sum_{k\geq 2}y_k-\mu_F F_1}x_i\sum_{\ell\neq i}x_\ell\times\\
\nonumber &\times e^{-s(\sum_k x_k+F_1)-t(\sum_k x_k+F_1)}=\\
\nonumber &=(N-1)\int ds dt \frac{1}{\mu_F+s+t}\left[\int dx~ x e^{-\mu x-(s+t)x}\right]^{2}\\
\nonumber &\times\left[\int dx~e^{-\mu x - (s+t)x}\right]^{N-2}\times \left[\int dy e^{-\mu y}\right]^{N-1}\\
\nonumber &=\frac{N-1}{\mu^{N-1}}\int_0^\infty ds dt \frac{1}{\mu_F+s+t}\frac{1}{(\mu+s+t)^{N+2}}=\\
&=\boxed{\frac{N-1}{\mu^{N-1}}L(N+2)}\ .
\end{align}

\begin{align}
\nonumber K_4 &= \int dx_1\cdots dx_N dF_1 dy_2\cdots dy_N ds dt\\
\nonumber &\times e^{-\mu \sum_k x_k -\mu\sum_{k\geq 2}y_k-\mu_F F_1}\sum_{k\geq 2}y_k\sum_{\ell\neq i}x_\ell\times\\
\nonumber &\times e^{-s(\sum_k x_k+F_1)-t(\sum_k x_k+F_1)}=\\
\nonumber &=(N-1)^2\int ds dt \frac{1}{\mu_F+s+t}\\
\nonumber &\times\left[\int dx~ x e^{-\mu x-(s+t)x}\right]\left[\int dx~e^{-\mu x - (s+t)x}\right]^{N-1}\\
\nonumber &\times \left[\int dy e^{-\mu y}\right]^{N-2}\int dy~y e^{-\mu y}\\
\nonumber &=\frac{(N-1)^2}{\mu^{N}}\int_0^\infty ds dt \frac{1}{\mu_F+s+t}\frac{1}{(\mu+s+t)^{N+1}}=\\
&=\boxed{\frac{(N-1)^2}{\mu^{N}}L(N+1)}\ .
\end{align}

\begin{figure}[h]
    \centering
    \fbox{\includegraphics[scale = 0.20]{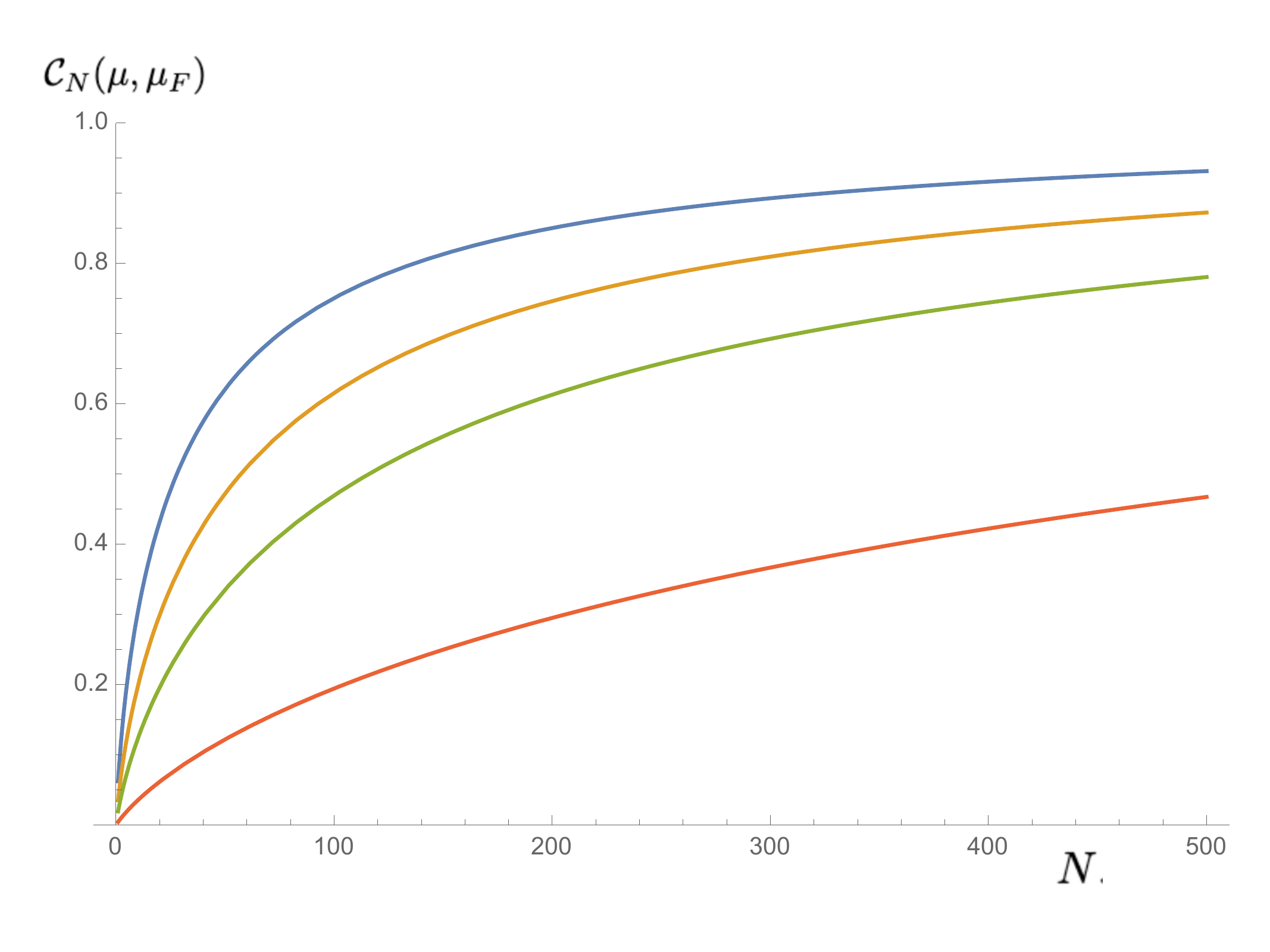}}
    \caption{Covariance $\mathcal{C}_N(\mu,\mu_F)$ between the approximate upstreamness and downstreamness for the random model (exact formula in Theorem \ref{thm1}). The parameters $(\mu,\mu_F)$ are $(1,0.1)$ (blue), $(2,0.1)$ (orange), $(2,0.05)$ (green), $(2,0.01)$ (red).}
    \label{fig:CNapp}
\end{figure}

Collecting all terms and simplifying, we arrive at the formula announced in Theorem \ref{thm1}. Plotting the covariance formula as a function of $N$ for different values of $\mu,\mu_F$ reveals that the covariance is always positive and increasing (see Fig. \ref{fig:CNapp}).

To prove the Corollary \ref{cor1}, we need to further compute $\mathbb{E}[r^{ 2}]$ and then simplify the resulting expression for the slope \eqref{eq:approximateslope}, yielding a slope $=+1$ for any $N$.

\begin{align}
\nonumber &\mathbb{E}[r^2] =\int_0^\infty \prod_{i=1}^N da_i p(a_i)\int_0^\infty dF~p_F(F)\left[\frac{\sum_k a_k}{\sum_k a_k + F}\right]^2\\
\nonumber &=\mu^N\mu_F \int d\bm a dF ds~s e^{-\mu\sum_k a_k-\mu_F F}\times \\
\nonumber &\left(\sum_k a_k\right)^2 e^{-s(\sum_k a_k+F)}\\
\nonumber &=\mu^N\mu_F \left[N\int d\bm a dF ds~s e^{-\mu\sum_k a_k-\mu_F F}a_1^2 e^{-s(\sum_k a_k+F)}+\right.\\
\nonumber &+\left. (N^2-N)\int d\bm a dF ds~s e^{-\mu\sum_k a_k-\mu_F F}a_1 a_2 e^{-s(\sum_k a_k+F)}\right]\\
\nonumber &=\mu^N\mu_F N \int_0^\infty ds~s \left[\int dx e^{-\mu x-s x}\right]^{N-1}\int dy~y^2 e^{-\mu y-s y}\times \\
\nonumber &\times\int dF e^{-\mu_F F-s F}+\mu^N\mu_F (N^2-N) \int_0^\infty ds~s\times\\
\nonumber &\times\left[\int dx e^{-\mu x-s x}\right]^{N-2}\left[\int dy~y e^{-\mu y-s y}\right]^2\int dF e^{-\mu_F F-s F}\\
\nonumber & =2\mu^N\mu_F N \int_0^\infty ds \frac{s}{(\mu+s)^{N+2}}\frac{1}{\mu_F+s}+\\
\nonumber &+\mu^N\mu_F (N^2-N) \int_0^\infty ds \frac{s}{(\mu+s)^{N+2}}\frac{1}{\mu_F+s}\\
\nonumber & =\boxed{\mu^N\mu_F(N^2+N)L(N+2)}\blacksquare
\end{align}

\end{document}